%% file: 0_main.tex
\documentclass[sigconf]{acmart}

\AtBeginDocument{%
  \providecommand\BibTeX{{%
    \normalfont B\kern-0.5em{\scshape i\kern-0.25em b}\kern-0.8em\TeX}}}

\copyrightyear{2022} 
\acmYear{2022} 
\setcopyright{acmcopyright}\acmConference[FAccT '22]{2022 ACM Conference on Fairness, Accountability, and Transparency}{June 21--24, 2022}{Seoul, Republic of Korea}
\acmBooktitle{2022 ACM Conference on Fairness, Accountability, and Transparency (FAccT '22), June 21--24, 2022, Seoul, Republic of Korea}
\acmPrice{15.00}
\acmDOI{10.1145/3531146.3533116}
\acmISBN{978-1-4503-9352-2/22/06}

\newcommand\enquote[1]{`#1'}
\usepackage{listings}
\usepackage{xcolor}
\usepackage{wrapfig}


\colorlet{punct}{red!60!black}
\definecolor{background}{HTML}{EEEEEE}
\definecolor{delim}{RGB}{20,105,176}
\colorlet{numb}{magenta!60!black}

\lstdefinelanguage{json}{
    basicstyle=\normalfont\ttfamily\tiny,
    showstringspaces=false,
    breaklines=true,
    frame=lines,
    backgroundcolor=\color{background}
}
\usepackage{caption,subcaption}

\newcommand*{\cm}{\checkmark}

\usepackage{color}

\begin{document}

\title{Goodbye Tracking? Impact of iOS App Tracking Transparency and Privacy Labels}


\author{Konrad Kollnig}
\email{konrad.kollnig@cs.ox.ac.uk}
\affiliation{%
  \institution{Department of Computer Science, University of Oxford}
  \streetaddress{Parks Road}
  \city{Oxford}
  \postcode{OX1 3QD}
  \country{United Kingdom}
}
\author{Anastasia Shuba}
\email{ashuba22@gmail.com}
\affiliation{%
  \institution{Independent Researcher}
  \country{USA}
}
\author{Max Van Kleek}
\email{max.van.kleek@cs.ox.ac.uk}
\affiliation{%
  \institution{Department of Computer Science, University of Oxford}
  \streetaddress{Parks Road}
  \city{Oxford}
  \postcode{OX1 3QD}
  \country{United Kingdom}
}
\author{Reuben Binns}
\email{reuben.binns@cs.ox.ac.uk}
\affiliation{%
  \institution{Department of Computer Science, University of Oxford}
  \streetaddress{Parks Road}
  \city{Oxford}
  \postcode{OX1 3QD}
  \country{United Kingdom}
}
\author{Nigel Shadbolt}
\email{nigel.shadbolt@cs.ox.ac.uk}
\affiliation{%
  \institution{Department of Computer Science, University of Oxford}
  \streetaddress{Parks Road}
  \city{Oxford}
  \postcode{OX1 3QD}
  \country{United Kingdom}
}

\renewcommand{\shortauthors}{Kollnig et al.}

\begin{abstract}
  \input{00_abstract}
\end{abstract}

\ccsdesc[500]{Security and privacy~Privacy protections}
\ccsdesc[300]{Security and privacy~Economics of security and privacy}
\ccsdesc[100]{Networks~Mobile and wireless security}

\keywords{mobile apps, Apple, iOS, data protection, privacy, platform policies, gatekeeper power, App Tracking Transparency, Privacy Nutrition Labels}

\maketitle

\section{Introduction}
\label{introduction}
\input{1_introduction}

\section{Background}
\label{sec:background}
\input{2_background}

\input{3_method}

\section{Results}
\label{sec:results}
\input{4_results}

\section{Discussion}
\label{sec:discussion}
\input{5_discussion}

\section{Conclusions \& Future Work}
\label{sec:conclusions}
\input{6_conclusions}

\begin{acks}
We thank Jake Stein and Alexander Fanta for helpful comments and Ulrik Lyngs for help with data analysis. Konrad Kollnig was funded by the UK Engineering and Physical Sciences Research Council (EPSRC) under grant number EP/R513295/1.
Max Van Kleek has been supported by the PETRAS National Centre of Excellence for IoT Systems Cybersecurity, which has been funded by the UK EPSRC under grant number EP/S035362/1. Max Van Kleek, Reuben Binns, and Nigel Shadbolt have been supported by the Oxford Martin School EWADA Programme.
\end{acks}

\bibliographystyle{ACM-Reference-Format}
\bibliography{references}


\appendix
\input{7_appendix}

\end{document}

%% file: 00_abstract.tex
Tracking is a highly privacy-invasive data collection practice that has been ubiquitous in mobile apps for many years due to its role in supporting advertising-based revenue models. 
In response, Apple introduced two significant changes with iOS 14: App Tracking Transparency (ATT), a mandatory opt-in system for enabling tracking on iOS, and Privacy Nutrition Labels, which disclose what kinds of data each app processes.
So far, the impact of these changes on individual privacy and control has not been well understood.
This paper addresses this gap by analysing two versions of 1,759 iOS apps from the UK App Store: one version from before iOS 14 and one that has been updated to comply with the new rules.

We find that Apple's new policies, as promised, prevent the collection of the Identifier for Advertisers (IDFA), an identifier for cross-app tracking.
Smaller data brokers that engage in invasive data practices will now face higher challenges in tracking users~--~a positive development for privacy.
However, the number of tracking libraries has~--~on average~--~roughly stayed the same in the studied apps.
Many apps still collect device information that can be used to track users at a group level (\textit{cohort tracking}) or identify individuals probabilistically (\textit{fingerprinting}).
We find real-world evidence of apps computing and agreeing on a fingerprinting-derived identifier through the use of server-side code, thereby violating Apple's policies. 
We find that Apple itself engages in some forms of tracking and exempts invasive data practices like first-party tracking and credit scoring from its new tracking rules.
We also find that the new Privacy Nutrition Labels are sometimes inaccurate and misleading, especially in less popular apps.

Overall, our observations suggest that, while Apple's changes make tracking individual users more difficult, they motivate a countermovement, and reinforce existing market power of gatekeeper companies with access to large troves of first-party data.
Making the privacy properties of apps transparent through large-scale analysis remains a difficult target for independent researchers, and a key obstacle to meaningful, accountable and verifiable privacy protections.

%% file: 1_introduction.tex
\begin{figure*}
    \begin{subfigure}{0.48\linewidth}
	    \centering
	    \includegraphics[width=0.6\linewidth]{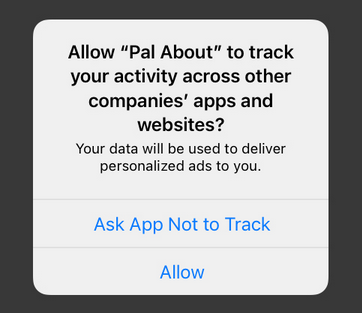}
		\caption{Apple Tracking Transparency (ATT).}
		\Description{This image shows the popup that is shown as part of Apple's new Tracking Transparency framework. Users can either \enquote{Allow} or \enquote{Ask App Not to Track}.}
		\label{fig:att}
	\end{subfigure}%
	\begin{subfigure}{0.48\linewidth}
	    \centering
	    \includegraphics[width=0.6\linewidth]{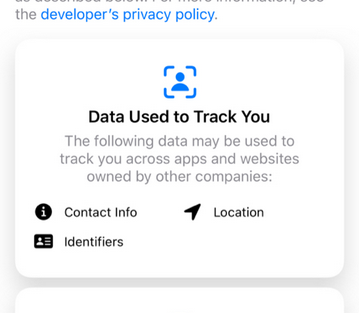}
		\caption{Privacy Nutrition Label.}
		\Description{This image shows an example of a privacy label, specifically types of \enquote{Data Used to Track You}.}
		\label{fig:privacy_label}
	\end{subfigure}
	\caption{Overview of Apple's new privacy measures, introduced with iOS 14~\cite{apple_tracking_definition}.}~\label{fig:apple_privacy}
	\Description{This figure shows an overview of Apple's new privacy measures.}
\end{figure*}

Tracking, the large-scale collection of data about user behaviour, is commonplace across both mobile app ecosystems, Android and iOS.
While some see tracking as a `necessary evil' to making apps available at lower prices by showing users personalised advertising and selling their data to third parties~\cite{anirudhchi2021,mhaidli_we_2019},
tracking can have highly disproportionate effects on the lives of individuals and society as a whole~\cite{van_kleek_better_2017,shklovski_leakiness_2014}.
As a countermeasure, Apple introduced the \textit{Apple Tracking Transparency} (ATT) framework~--~alongside mandatory \textit{Privacy Nutrition Labels}~\cite{kelley_nutrition_2009,10.1145/1753326.1753561}~--~with iOS 14, see Figure~\ref{fig:apple_privacy}.

The emergence of more robust privacy measures in everyday technology is partly motivated by new data protection and privacy laws around the globe, particularly the General Data Protection Regulation (GDPR) in the EU and UK since May 2018~\cite{kollnig_before_2021}.
Among other aspects, the GDPR protects any data that can be related to individuals (\enquote{personal data}), and requires a legal basis for any processing of such personal data.
This requirement has the effect that app tracking, which usually classifies as \enquote{high-risk} data processing, needs prior user consent~\cite{apps_consent_2021,kollnig_2021}.
Additionally, the 2009 ePrivacy Directive, which regulates data processing in electronic systems in the EU and UK, also requires consent to tracking~\cite{kollnig_2021,datenschutzkonferenz_orientierungshilfe_2021}.
Despite these legal requirements, a large proportion of apps engaging in tracking have in the past been observed not to seek the required prior user consent~\cite{apps_consent_2021,kollnig_2021,reyes_wont_2018}.

Starting with iOS 14.5 in April 2021, iOS apps must now ask users for explicit permission before tracking them, see Figure~\ref{fig:att}.
If an iOS user asks an app not to track, this has the direct effect that this app cannot access the Identifier for Advertisers (IDFA) anymore.
The IDFA is a random, unique identifier provided by the operating system to apps for tracking users across multiple sessions of a single app and across apps.
Additionally, apps are obliged to stop certain tracking practices under the Apple's App Store policies (more in Section~\ref{sec:discussion}).
Preliminary data suggests that the vast majority of users (between 60\% and 95\%) choose to refuse tracking when asked for it under the new system~\cite{att_optout1,att_optout2,att_optout3}.

While potentially good for user privacy, the ATT has been reported to have vastly increased Apple’s share of advertising on iOS~--~as part of its Apple Search Ads on the App Store~--~and to have decreased the efficacy of ads from competing companies.
An important reason for this, as argued by Eric Seufert and others, is that Apple's own tracking technologies may not fall under Apple's definition of tracking~\cite{apple_selfpreference}.
It has also been reported that many marketing companies have shifted advertising budgets from iOS to Android~\cite{att_marketingbudgets}.
The Financial Times estimated that the loss for leading tech companies from the new policy would be around \$10bn~\cite{att_impact_revenues2}, but also reported that companies deemed the \enquote{effect of Apple’s privacy changes was less than feared}~\cite{att_impact_revenues}.
Apple's privacy changes may prompt a rise in paid apps and in-app purchases~\cite{kesler_att_2022}, and thereby particularly affect those individuals who are already worse off financially.


In addition to the changes relating to the ATT, app developers must now self-declare what types of data they collect from users, and for what purposes~--~called \textit{Privacy Nutrition Labels}~\cite{kelley_nutrition_2009,10.1145/1753326.1753561}, see Figure~\ref{fig:apple_privacy}.
As such, these labels aim to make it easier for end-users to understand the data practices of apps, instead of having to study lengthy privacy policies, which few users do~\cite{mcdonald_cost_2008}.
There is, however, a risk that many users may just ignore the new (and potentially oversimplified) privacy labels (as they commonly do with privacy policies~\cite{mcdonald_cost_2008}), gain a false sense of security, or not understand the consequences for their privacy (which tends to be highly individual~\cite{nissenbaum_privacy_2004}), and that developers may not honestly self-declare their actual data practices~\cite{nutrition_labels}.
Despite these concerns, the labels have the potential to shift developers' existing data practices towards being more privacy-preserving, through increased transparency and end-user awareness.

Based on the above observations, this paper analyses the following research questions:
\begin{enumerate}
    \item What impact have the ATT and Privacy Nutrition Labels had~--~thus far~--~on tracking, particularly on the extent and quality of tracking?
    \item To what extent do apps disclose their tracking practices in their Privacy Nutrition Labels?
    \item What implications do the ATT and Privacy Nutrition Labels have for the power relations between the actors in the digital advertising system, including mobile OS providers, digital advertisers, app developers and marketers?
\end{enumerate}

To analyse these questions, this paper analyses privacy in 1,759 iOS apps,
for each of which we downloaded two versions: one from before Apple's new rules and one that has been updated since.
We use a combination of app code and network analysis to gain rich insights into the data practices of the studied apps.

The remainder of this paper is structured as follows. We first review related work in Section~\ref{sec:background}.
Next, we introduce our app download and analysis methodology in Section~\ref{sec:methodology}.
We turn to the results from our app code and network analysis in Section~\ref{sec:results}.
We discuss our findings in Section~\ref{sec:discussion} and the limitations of our study in Section~\ref{sec:limitations}.
We conclude the paper and outline direction for future work in Section~\ref{sec:conclusions}.
Code and data to replicate our results are available at \url{https://www.platformcontrol.org/}.

%% file: 2_background.tex
\subsection{Related work}
\label{sec:related-work}
Previous research extensively studied privacy in mobile apps. Two main methods have emerged in the academic literature: dynamic and static analysis.

\emph{Dynamic analysis} observes the run-time behaviour of an app, to gather evidence of sensitive data leaving the device. Early research focused on OS instrumentation, i.e. modifying Android~\cite{enck_taintdroid_2010} or iOS~\cite{agarwal_protectmyprivacy_2013}. With growing complexity of mobile operating systems, recent work has shifted to analysing network traffic~\cite{privacyguard_vpn_2015,nomoads_2018,free_v_paid_2019,reyes_wont_2018,van_kleek_better_2017,ren_recon_2016,nomoads_2018,shuba_nomoats_2020}.
This comes with certain limitations.
One problem is limited scalability, since every app is executed individually. 
Another issue is that not all privacy-relevant parts of apps may be invoked during analysis, potentially leading to incomplete results.

\emph{Static analysis} dissects apps without execution. Usually, apps are decompiled, and the obtained program code is analysed~\cite{han_comparing_2013,pios_2011}.
The key benefit of static analysis is that it can analyse apps quickly, allowing it to scale to millions of apps~\cite{china_2018,playdrone_2014,binns_third_2018,chen_following_2016,kollnig_before_2021}.
However, static analysis can involve substantial computational effort and~--~unlike dynamic analysis~--~does not allow the observation of real data flows because apps are never actually run.
Programming techniques, such as the use of code obfuscation and native code, can pose further obstacles.
This is especially true for iOS apps, which are often harder to analyse and decompile~--~compared to Android~--~and are encrypted by default~\cite{kollnig2021iphones,maps_2019,binns_third_2018}.
While this iOS encryption might legitimately protect \textit{paid} apps against piracy, Apple also encrypts all free apps downloaded from the App Store.
By contrast, Google only encrypts paid apps (not free ones) when downloaded from its Play Store.
The encryption of iOS apps by Apple~--~even of free ones~--~is problematic for research efforts because it drives researchers into legal grey areas of copyright law~\cite{kollnig2021iphones}.
Partly because of these difficulties, our recent work~\cite{kollnig2021iphones} was the first large-scale app privacy analysis study on iOS apps since 2013~\cite{agarwal_protectmyprivacy_2013}.
We avoided legal problems relating to copyright law by conducting part of the analysis on-device through using the popular app instrumentation tool \texttt{Frida}~\cite{frida}.

In this paper, we follow the methodology of our previous paper, which used a combination of both dynamic and static analysis, so as to compare the privacy practices of the studied apps before and after the introduction of Apple's new privacy rules.
We discuss our methodology for this paper in more detail in Section~\ref{sec:methodology}.

\subsection{Regulation of App Platforms}
\label{sec:regulation}

The centrality of app platforms~--~i.e. Apple's iOS and Google's Android ecosystem~--~makes them a target for effective privacy regulation, however such regulation is limited ~\cite{o_fathaigh_european_2019,hoboken2021}. The US Federal Trade Commission (FTC) established some baseline rules for app stores in 2013. They strongly recommended to app platforms to require just-in-time consent for sensitive data access, to seek privacy policies from app developers, and to implement system-wide opt-out mechanism from data collection~\cite{ftc_app_stores}. Despite not being law, Google and Apple followed many of the recommendations, and have not seen further public recommendations from the FTC since.

In the EU and UK, there exists no targeted regulation of app stores. The Regulation on platform-to-business relations (P2BR) contains general provisions for online intermediaries, including app stores, but does little to enact better privacy protections \cite{o_fathaigh_european_2019}.
Data protection laws, such as the GDPR and the ePrivacy Directive,
arguably place the primary responsibility for data protection with the app developers, not usually with app platform providers~--~although this is subject to ongoing debate; this lack of data protection obligations within the entire software development process -- not just deployment -- has been widely criticised \cite{bygrave_data_2017,jasmontaite_data_2018}.

While no targeted regulation exists, app platforms face increasing scrutiny by courts and regulators.
In the case \textit{Epic Games v Apple} running since 2020, a US District Court judge largely found no monopolistic behaviour of Apple, but did identify some anticompetitive conduct in Apple's business practices.
The judge ordered Apple to allow app developers to inform app users of alternative payment methods.
Both Apple and Epic Games have appealed the ruling.
In the EU, following a complaint of Spotify against Apple from 2019, the European Commission identified multiple anticompetitive aspects about Apple's ecosystem in a preliminary ruling~--~the case is, however, still ongoing.
In January 2022, the Dutch competition authority demanded changes from Apple to its App Store policies; Apple has to date not fulfilled the demands of the regulators in their entirety, and has instead chosen to pay a weekly penalty of €5 million up to a maximum of €50 million~\cite{dutch_authority}.

The challenges in keeping up with regulation of platforms have spurred a recent countermovement by lawmakers.
In South Korea, parliament amended the Telecommunication Business Act to force app stores to allow alternative payment methods and reduce commissions~\cite{korea}.
In response, Apple lowered the share it takes from App Store revenues of small developers (making less than \$1 million per year) from 30\% to 15\%.
In the US, Congress is debating a new Open App Markets Act that aims to address common competition concerns around app stores and passed the Senate Judiciary Committee with a strong a 20–-2 bipartisan vote in February 2022.
In the EU, lawmakers are seeking to enact two new pieces of legislation that aim to improve the regulation of digital markets, the Digital Markets Act and the Digital Services Act.
Any new legal requirement for app platforms will likely have implications worldwide, due to the nature of digital ecosystems.

In sum, there currently exist few specific legal obligations for app platforms. Instead, they are encouraged to self-regulate their conduct.
The following analysis shall shine a light on how the recent policy changes by Apple, a highly prominent example of this self-regulation,
have affected the actual privacy practices of mobile apps.

%% file: 3_method.tex
\section{Methodology}
\label{sec:methodology}

\begin{figure}
  \centering
  \includegraphics[width=0.95\linewidth]{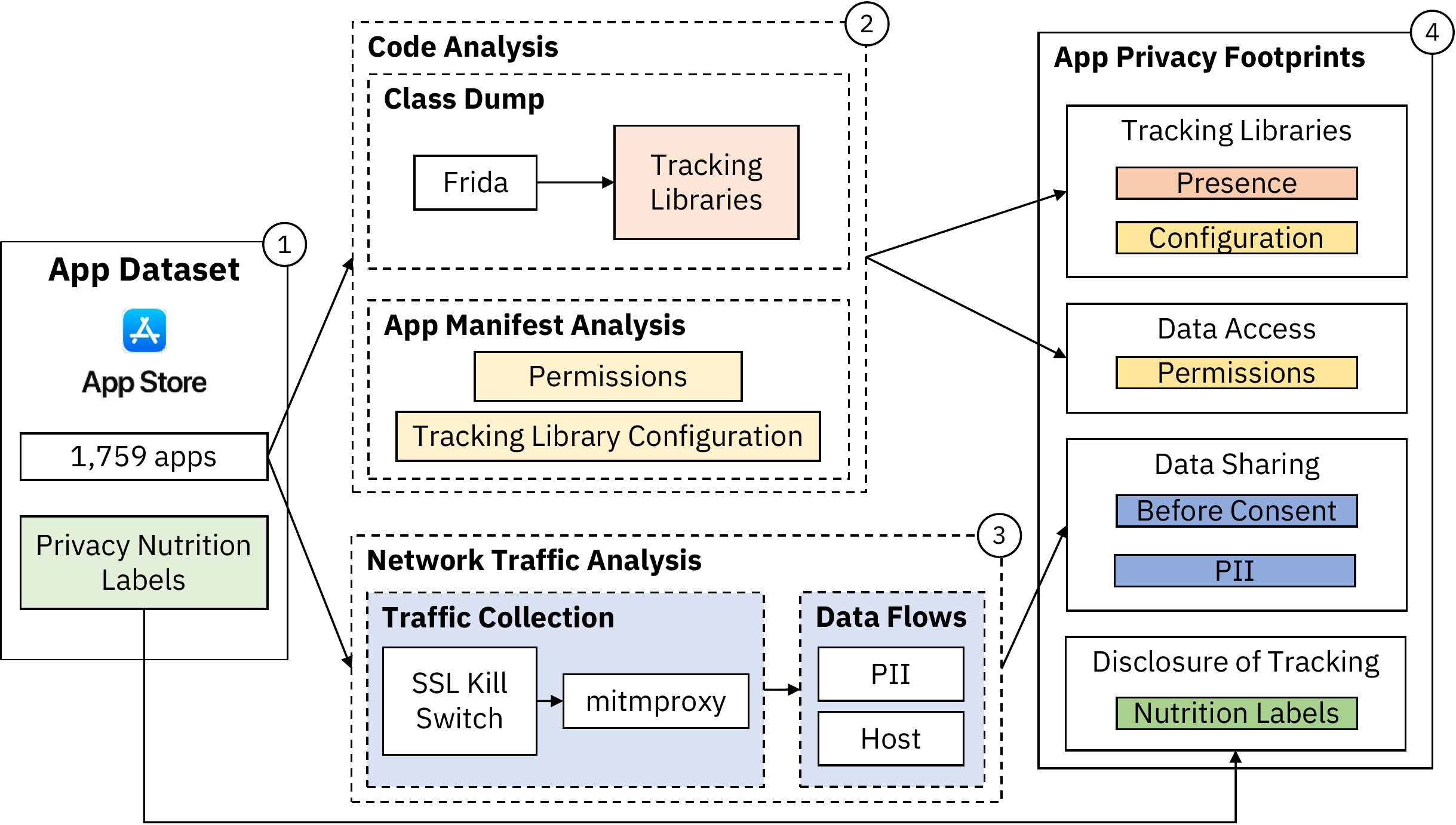}
  \caption{Overview of our analysis methodology (Section \ref{sec:methodology}): First, (1) we select and download 1,759 apps from before the introduction of the ATT, and 1,759 from after. We also collect apps' Privacy Nutrition Labels.
    Next, we perform (2) \textbf{Code Analysis} to examine permissions and tracking libraries usage; and
    (3) \textbf{Network Traffic Analysis} to analyse tracking domains contacted at the first app start and the sharing of personal data.
    The results of this analysis (Section \ref{sec:results}) are detailed \textbf{App Privacy Footprints} (4) of the downloaded apps.}
    \label{fig:method_flow}
    \Description{This figures shows a visualisation of the methodology that is also explained in Section~\ref{sec:methodology}.}
\end{figure}

In this section, we describe our analysis methodology (depicted in Figure \ref{fig:method_flow}),
which
follows the one that we previously used for a comparative analysis of iOS and Android apps' privacy practices~\cite{kollnig2021iphones}.
Code and data to replicate our results are available at \url{https://www.platformcontrol.org/}.
We therefore keep our description of the methodology short and refer the reader to the original paper for details.

\subsection{App Selection and Download}

This section details our process for selecting and downloading apps from the Apple App Store (step 1 in Figure~\ref{fig:method_flow}).
For the selection of apps, we revisited the same 12,000 iOS apps as in our previous study~\cite{kollnig2021iphones}.
These apps were selected by first generating a large list of apps available on the Apple App Store between December 2019 and February 2020.
We then downloaded a random subset ($n=12,000$) of those apps that were last updated since 2018 so as to focus on apps currently in use.
For this work, we re-downloaded those apps that were updated to comply with Apple's ATT and privacy label rules, in October 2021.
This resulted in a dataset of 1,759 \textit{pairs} of apps, one from before iOS 14 and one from after.
This number of apps is comparatively small because many apps had not yet been updated since the new rules, while some other apps had been removed from the store (2,713 out of 12,000 apps were not available on the App Store anymore).
We additionally scraped the Privacy Nutrition Labels for the newly downloaded apps.

\subsection{Code Analysis}
To identify the presence of tracking libraries (step 2 in Figure~\ref{fig:method_flow}), we extracted the names of all classes loaded by each app using the tool \texttt{Frida}~\cite{frida} and checked them against a list of known tracker class names from our previous paper~\cite{kollnig2021iphones}. We also examined the app manifest (every iOS app must provide such a file) to determine how certain tracking libraries are configured -- many tracking libraries allow developers to restrict data collection using settings in the manifest file, e.g. to disable the collection of unique identifiers or the automatic SDK initialisation at the first app start.
This can help set up tracking libraries in a legally compliant manner.
For example,
`Data minimisation' is one of the key principles of GDPR (Article 5.1 (c)), and user opt-in is required prior to app tracking in the EU and UK~\cite{kollnig_2021}.
We analysed the privacy settings provided by some of the most prominent tracking libraries: Google AdMob, Facebook, and Google Firebase.

Beyond analysing tracking in apps, we also obtained a list of permissions that apps can request.
Permissions form an important part of the security model of iOS as they protect sensitive information on the device, such as apps' access to the camera or address book.
As such, permissions are different to the new privacy labels, which do not affect the runtime behaviour of apps.
We extracted apps' permissions by automatically inspecting the manifest file.

\subsection{Network Analysis}

To analyse apps's network traffic (step 3 in Figure~\ref{fig:method_flow}), we executed every app on a real device~--~one iPhone SE 1st Gen with iOS 14.2, and one with iOS 14.8~--~for 30 seconds without user interaction. We captured network traffic using the tool \texttt{mitmdump}.
We disabled certificate validation using \texttt{SSL Kill Switch 2}, after gaining system-level access on both iPhones (known as `jailbreak').
On the iPhone with iOS 14.2, we did not opt-out from ad personalisation from the system settings, thereby assuming user opt-in to use the IDFA (reflecting the assumption that many users, who would reject tracking, do not do so because the option is in the less prominent settings on the OS~\cite{kollnig2021iphones}).
On the iPhone with iOS 14.8, we asked all apps not to track from the system settings.
Although in Android privacy research real user behaviour is simulated via various automation tools~\cite{van_kleek_x-ray_2018,ren_recon_2016,okoyomon_ridiculousness_2019,han_price_2020,binns_measuring_2018,reyes_wont_2018,shuba_nomoats_2020}, Apple's restrictions on debugging and instrumentation have hindered the development of such tools for iOS.
Tracking libraries are usually initialised at the first app start and without user consent~\cite{kollnig_2021, reyes_wont_2018,nguyen_share_first_consent_2021,kollnig2021iphones}, and they can thus be detected without user interaction in the network traffic, as done in our analysis.

%% file: 4_results.tex
\begin{figure*}
    \begin{subfigure}{0.48\linewidth}
	    \centering
	    \includegraphics[width=\linewidth]{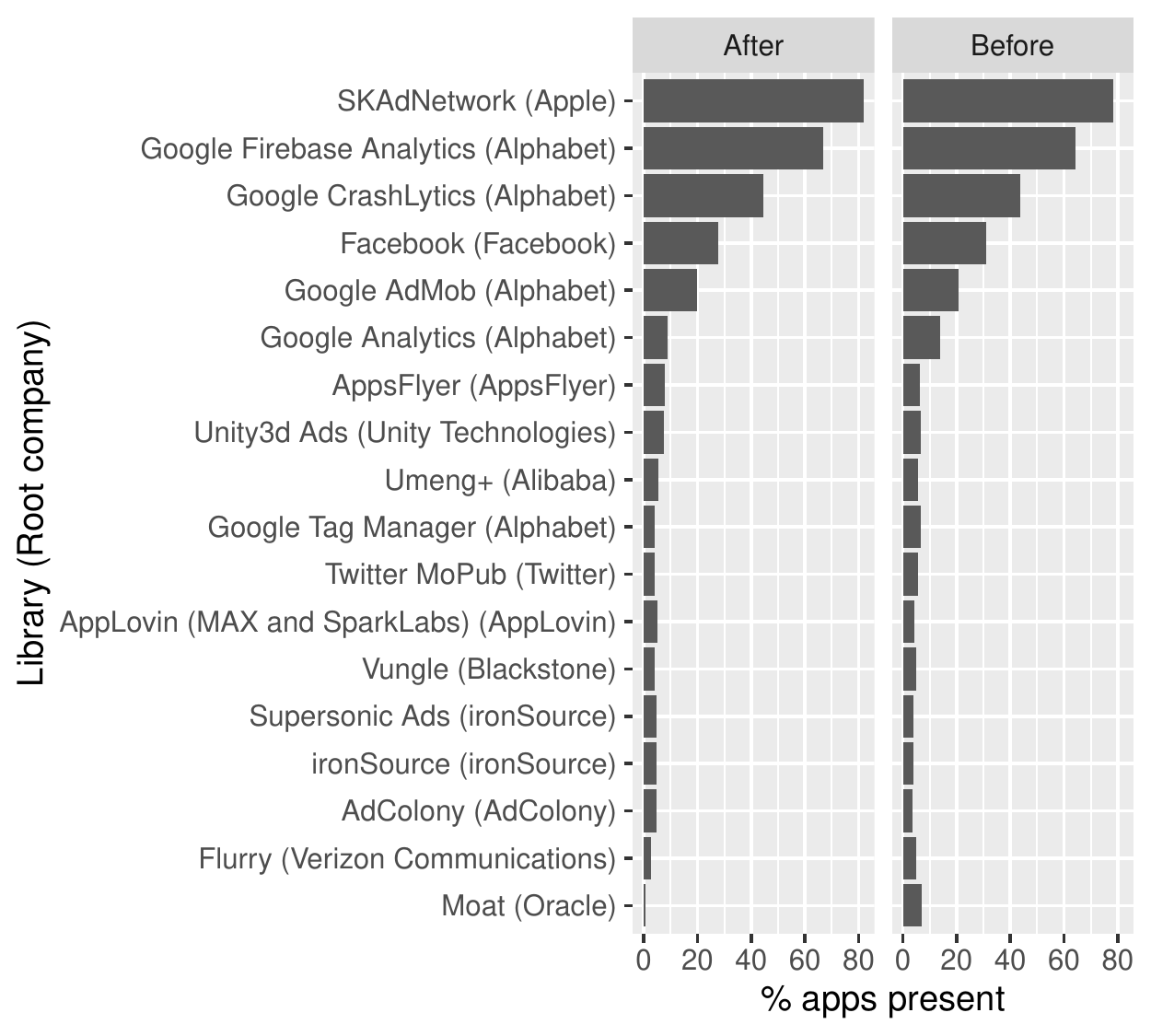}
	    \footnotesize
	    \begin{tabular}{lrrrrrr} \toprule
			& Median
			& Mean
			& Q1
			& Q3
			& Count $>10$
			& None \\
			\midrule
			Before & 3 & 3.7 & 2 & 5 & 4.75\% & 13.61\% \\
			After  & 3 & 3.6 & 2 & 4 & 4.75\% & 12.48\% \\ \bottomrule
		\end{tabular}
		\caption{Top tracking libraries in app code.}
	\end{subfigure}
	\begin{subfigure}{0.48\linewidth}
	    \centering
	    \includegraphics[width=\linewidth]{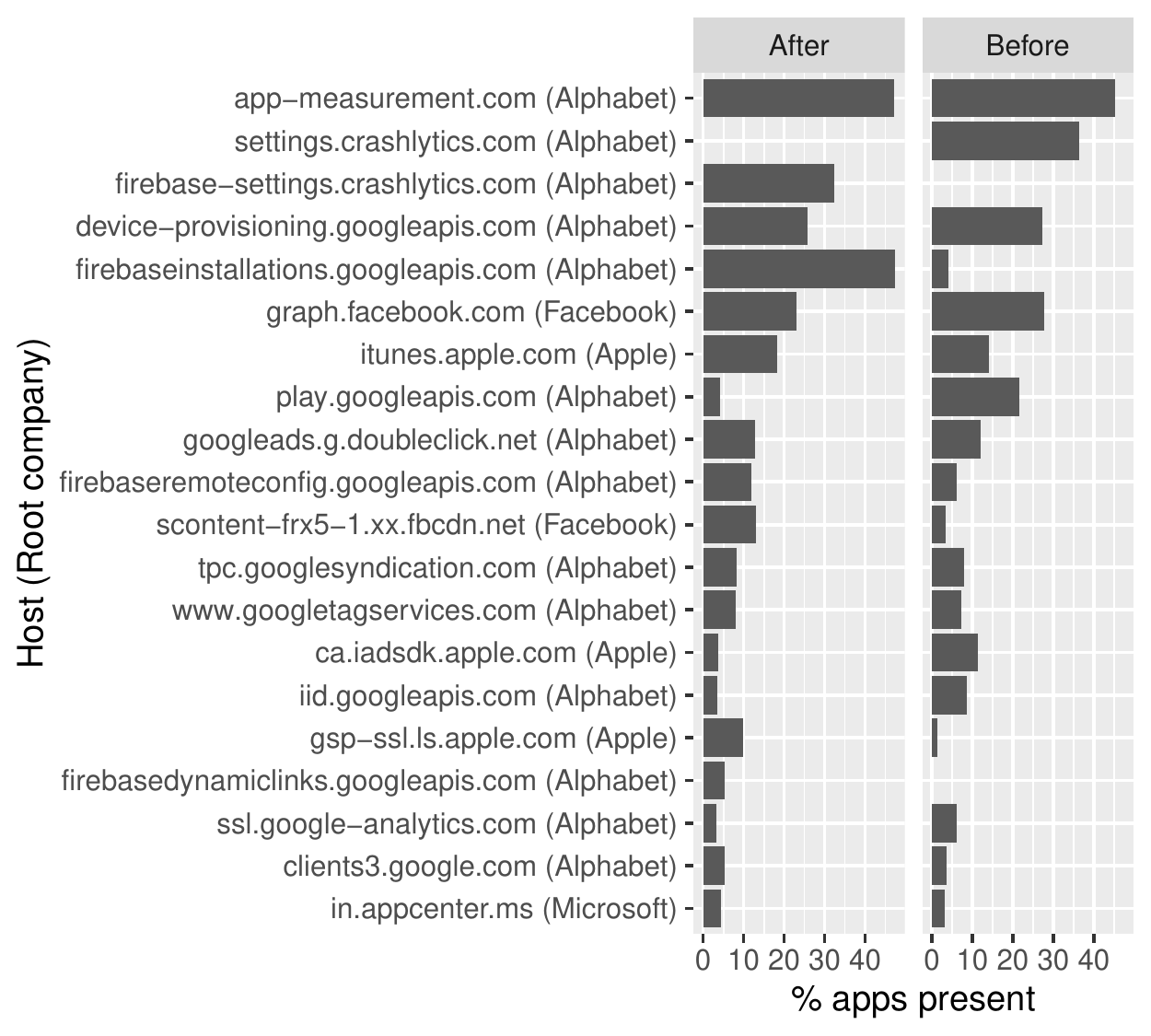}
	    \footnotesize
	    \begin{tabular}{lrrrrrr} \toprule
			& Median
			& Mean
			& Q1
			& Q3
			& Count $>10$
			& None \\
			\midrule
			Before & 3 & 4.0 & 1 & 6 & 4.75\% & 13.43\% \\
			After  & 4 & 4.7 & 2 & 7 & 7.19\% & 10.70\% \\ \bottomrule
		\end{tabular}
		\caption{Top tracking hosts contacted at the first app start.}
	\end{subfigure}
	\caption{Third-party libraries (integrated in apps, but not necessarily activated) and contacted tracking domains of apps, as well as the companies owning them (in brackets). Shown are the top 15 tracking libraries and domains for before and after the new privacy changes under iOS 14. 
	}~\label{fig:apps_trackers}
	\Description{This figure shows the analysis results for tracking libraries in apps and contacted tracking hosts at the first app start, each both as a bar chart and a table.}
\end{figure*}

In this section, we present our findings from analysing two versions~--~one from before and one from after the release of iOS 14 and the ATT~--~of 1,759 iOS apps (step 4 in Figure~\ref{fig:method_flow}).
We analysed 199.6~GB of downloaded apps, extracted 3.2~GB in information about classes in apps' code, and collected 3.9~GB of data in apps' network traffic.
Installing and instrumentation failed for 74 iOS apps; we have excluded these apps from our subsequent analysis and focus on the remaining 1,685 apps.

First, we focus on the tracking libraries found in the code analysis (Section~\ref{sec:static_tracking}) and whether or not they were configured for data minimisation (Section~\ref{sec:static_tracking_config}).
Next, in Section~\ref{sec:data_access}, we analyse apps' access to the IDFA (which is now protected by the ATT) and also their permissions.
Following up, in Section~\ref{sec:data_sharing}, we report on the data sharing of apps before consent is provided, with a particular focus on whether apps that are instructed not to track actually do so in practice.
Lastly, in Section~\ref{sec:nutrition_labels}, we check whether and to what extent apps disclose their tracking practices in their Privacy Nutrition Labels.

\subsection{Tracking Libraries}
\label{sec:static_tracking}

Apps from both before the ATT and after widely used tracking libraries
(see Figure~\ref{fig:apps_trackers}a).
The median number of tracking libraries included in an app was 3 in both datasets.
The mean before was 3.7, the mean after was 3.6.
4.75\% of apps from before ATT contained more than 10 tracking libraries, compared to 4.75\% after. 86.39\% contained at least one before ATT, and 87.52\% after.

The most prominent libraries have not changed since the introduction of ATT.
The top one was the SKAdNetwork library (in 78.4\% of apps before, and 81.8\% after). While part of Apple's privacy-preserving advertising attribution system, this library discloses information about what ads a user clicked on to Apple, from which Apple could (theoretically) build user profiles for its own advertising system.
Following up with Apple about this potential issue (by one of the authors exercising the GDPR's \textit{right to be informed} under Article 13),
they did not deny the fact that this data might be used for advertising, but assured us that any targeted ads would only be served to segments of users (of at least 5,000 individuals with similar interests).
Google Firebase Analytics ranked second (64.3\% of apps from before ATT, and 67.0\% after), and Google Crashlytics third (43.6\% before, 44.4\% after).

Overall, Apple's privacy measures seem not to have affected the integration of tracker libraries into \textit{existing} apps.

\subsubsection{Configuration for Data Minimisation}
\label{sec:static_tracking_config}

Among the apps that used Google AdMob, 2.9\% of apps from before and 4.5\% from after chose to delay data collection.
Choosing to delay data collection can be helpful for app developers, to seek consent before enabling tracking and to fulfil legal obligations.
Among the apps using the Facebook SDK, there was an increase in those which delayed the sending of app events (6.7\% before, and 12.5\% after); an increase in those which delayed the SDK initialisation (1.0\% before ATT, 2.2\% after), and an increase in those which disabled the collection of the IDFA (5.0\% before, 8.6\% after). Among apps using Google Firebase,
0.6\% permanently deactivated analytics before ATT and 0.8\% after, 0.0\% disabled the collection of the IDFA before and 0.6\% after, and 0.6\% delayed the Firebase data collection before ATT and 1.0\% after.

Overall, we found that only a small fraction of apps made use of data-minimising SDK settings in their manifest files.
One reason for this observation might be that some developers are not aware of these settings because tracking companies tend to have an interest in less privacy-preserving defaults regarding data collection~\cite{mhaidli_we_2019,kollnig_2021}.
This fraction has subtly increased since the introduction of the ATT.

\subsection{Data Access and Permissions}
\label{sec:data_access}

\begin{figure}
    \centering
    \includegraphics[width=0.95\linewidth]{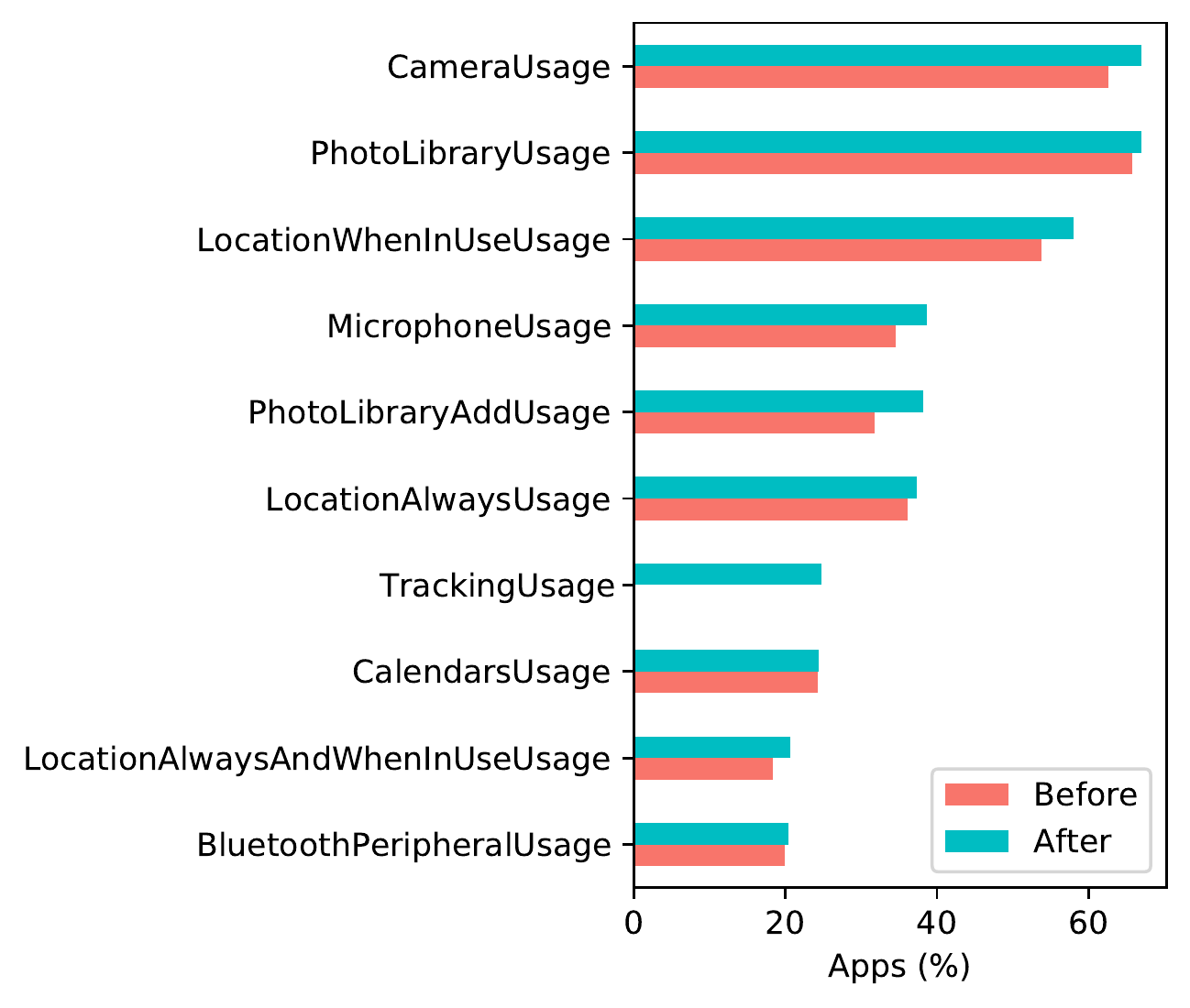}
    \caption{Top 10 permissions that apps can request.}
    \label{fig:permissions}
    \Description{This figure shows a bar chart of the top 10 permissions.}
\end{figure}

\textbf{Most prevalent permissions.}
Figure~\ref{fig:permissions} shows the most prevalent permissions before and after the introduction of the ATT.
On average, there was an increase in permission use ($4.3$ permissions before, $4.7$ after~--~excluding the new \textit{Tracking} permission).
\texttt{CameraUsage} (for camera access) was the most common permission (62.6\% before ATT, 66.9\% after), closely followed by
\texttt{PhotoLibraryUsage} (65.8\% before ATT, 66.9\% after), and \texttt{LocationWhenInUseUsage} (53.8\% before ATT, 58.0\% after).

\textbf{Tracking permission and access to IDFA.}
As part of ATT, apps that want to access the IDFA or conduct tracking must declare the \texttt{TrackingUsage} permission in their manifest.
24.7\% of apps from our dataset chose to declare this permission, and might ask users for tracking.
At the same time, the share of apps that contain the \texttt{AdSupport} library, necessary to access the IDFA in the app code, stayed unchanged at 50.8\% of apps.
This means that 50.8\% of apps from after the ATT could access the IDFA on earlier versions of iOS than 14.5, but only 24.7\% can on iOS 14.5 or higher.

\textbf{Tracking permission and integration of tracking SDKs.}
The share of apps that both contained a tracking library and could request tracking varied somewhat between the used tracking library.
69.3\% of the 350 apps that integrated Google AdMob declared the \texttt{TrackingUsage} permission;
78.7\% of the 110 apps that integrated Unity3d Ads;
50.0\% of the 116 apps that integrated Moat; and 77.3\% of the 54 apps that integrated Inmobi.
Whether the app is from before or after the ATT, the vast majority of apps (between 97 and 100\%) that integrated any of these tracking libraries also integrated the \texttt{AdSupport} library, and could therefore access the IDFA if running on iOS versions before 14.5.

\subsection{Data Sharing}
\label{sec:data_sharing}

\subsubsection{Before Consent}
\label{sec:data_sharing_consent}
This section analyses how many tracking domains apps contacted before any user interaction has taken place; the next Section~\ref{sec:data_sharing_pii} then analyses what data was shared with trackers.
Since tracking libraries usually start sending data right at the first app start~\cite{kollnig_2021,reyes_wont_2018,nguyen_share_first_consent_2021,kollnig2021iphones}, this approach provides additional evidence as to the nature of tracking in apps~--~and without consent.
Our results are shown in Figure~\ref{fig:apps_trackers}b.

The average number of tracking domains contacted was somewhat higher for apps from after the introduction of the ATT (4.0 before, 4.7 after).
The most popular domains were related to Google's analytics services: \path{firebaseinstallations.googleapis.com} (4.1\% of apps before the ATT, 47.4\% after) and \path{app-measurement.com} (45.2\% before, 47.2\% after).
Since both endpoints are related to Google Firebase, the large increase in \path{firebaseinstallations.googleapis.com} prevalence likely reflects internal restructuring of Firebase following Google's acquisitions of other advertising and analytics companies.
For example, Google acquired the crash reporting software Crashlytics from Twitter in January 2017, which is clearly reflected in our data. Google deprecated the old API endpoint (\path{settings.crashlytics.com} and changed it to \path{firebase-settings.crashlytics.com}) from November 2020.
This had the direct effect that all Crashlytics users must now also use Google Firebase.
The domain \path{settings.crashlytics.com} was contacted by 36.4\% for apps from before the ATT, and \path{firebase-settings.crashlytics.com} by 32.3\% after the ATT.
While this might point to a small difference in the adoption of Google Crashlytics, the exact same number of apps (734, 43.6\%) integrated the Crashlytics library into their code, before and after the ATT.
Similarly, the exact same number of apps integrate the Facebook SDK (523, 31.1\%); the share of apps that contacted the associated API endpoint \path{graph.facebook.com} at the first start fell from 27.7\% to 23.1\%.
The Google Admob SDK, too, was integrated in the same number of apps (350, 20.8\%), and did not see a decline in apps that contact the associated API endpoint \path{googleads.g.doubleclick.net} (12.1\% before, 12.9\% after).

Overall, data sharing with tracker companies before any user interaction remains common, even after the introduction of the ATT. This is in potential violation with applicable data protection and privacy laws in the EU and UK, which require prior consent~\cite{kollnig_2021}.

\subsubsection{Exposure of Personal Data}
\label{sec:data_sharing_pii}

We found that 26.0\% of apps from before the ATT shared the IDFA over the Internet, but none from after the ATT.
In this sense, the ATT effectively prevents apps from accessing the IDFA.
Despite Apple's promises, closer inspection of the network traffic showed that both Apple and other third parties are still able to engage in user tracking.

\begin{figure}
    \centering
	\small
    \begin{tabular}{llrr} \toprule
    Information      & Example             & Before     & After \\ \midrule
    iPhone Name      & MyPhone             & 2.5\%      & 4.2\%     \\
    iPhone Model     & iPhone8,4$\mid$iPhone SE & 60.2\%& 74.5\%    \\
    Carrier          & Three               & 20.2\%     & 20.2\%    \\
    Locale           & en\_GB$\mid$en-gb       & 85.7\%     & 90.1\%    \\
    CPU Architecture & ARM64$\mid$16777228 & 13.7\%     & 16.1\%    \\
    Board Config     & N69uAP              & 3.1\%      & 4.5\%     \\
    OS Version       & 14.8$\mid$18H17     & 79.9\%     & 86.9\%    \\
    Timezone         & Europe/London       & 3.9\%      & 3.4\%     \\ \bottomrule
    \end{tabular}
    \caption{Proportion of \textit{all} apps that shared device information. This information can potentially be used for fingerprinting or cohort tracking.}
    \label{tab:pii}
    \Description{This figure shows a table with types of data that have been shared by apps, as well as the prevalence of such sharing across all studied apps.}
\end{figure}

We found that iPhones continued to share a range of information with third-parties, that can potentially be used for device fingerprinting or cohort tracking, see Table~\ref{tab:pii}.
Only \textit{timezone} saw a subtle decrease in the number of apps that shared this information.
It is not clear why apps need to access or share some of this information, e.g. the carrier name (shared by 20.2\% of apps) or the iPhone name (shared by 3--4\% of apps).
Meanwhile, some types of information, particularly the iPhone name, might allow the identification of individuals, especially when combined with other information.

\begin{table*}
    \centering
    \footnotesize
    \begin{tabular}{llrcccc}
    \toprule
    Domain                                        & Company  &   Apps & User ID & Locale & Model & OS Version \\ \midrule
    \texttt{firebaseinstallations.googleapis.com} & Google   & 47.4\% & \cm     & \cm    &       &            \\
    \texttt{app-measurement.com}                  & Google   & 47.2\% & \cm     & \cm    &       &            \\
    \texttt{firebase-settings.crashlytics.com}    & Google   & 32.3\% & \cm     & \cm    & \cm   & \cm        \\
    \texttt{device-provisioning.googleapis.com}   & Google   & 25.8\% & \cm     & \cm    & \cm   & \cm        \\
    \texttt{graph.facebook.com}                   & Facebook & 23.1\% & \cm     & \cm    & \cm   & \cm        \\
    \texttt{itunes.apple.com}                     & Apple    & 18.3\% & \cm     & \cm    & \cm   & \cm        \\
    \texttt{fbcdn.net}                            & Facebook & 13.0\% &         & \cm    &       &            \\
    \texttt{googleads.g.doubleclick.net}          & Google   & 12.9\% & \cm     & \cm    & \cm   & \cm        \\
    \texttt{firebaseremoteconfig.googleapis.com}  & Google   & 11.8\% & \cm     & \cm    &       &            \\
    \texttt{gsp-ssl.ls.apple.com}                 & Apple    &  9.9\% & \cm     & \cm    & \cm   & \cm        \\
    \texttt{tpc.googlesyndication.com}            & Google   &  8.3\% &         & \cm    &       & \cm        \\
    \texttt{www.googletagservices.com}            & Google   &  8.1\% &         & \cm    &       & \cm        \\
    \texttt{clients3.google.com}                  & Google   &  5.3\% &         & \cm    &       &            \\
    \texttt{firebasedynamiclinks.googleapis.com}  & Google   &  5.2\% & \cm     & \cm    &       & \cm        \\
    \texttt{in.appcenter.ms}                      & Microsoft&  4.3\% & \cm     & \cm    & \cm   & \cm        \\ 
    \texttt{play.googleapis.com}                  & Google   &  4.2\% & \cm     & \cm    & \cm   & \cm        \\
    \texttt{skadsdk.appsflyer.com}                & AppsFlyer&  4.0\% & \cm     & \cm    &       &            \\ 
    \texttt{gsp64-ssl.ls.apple.com}               & Apple    &  3.9\% &         & \cm    & \cm   & \cm        \\
    \texttt{api.onesignal.com}                    & OneSignal&  3.7\% &         & \cm    &       &            \\ 
    \texttt{ca.iadsdk.apple.com}                  & Apple    &  3.7\% & \cm     & \cm    & \cm   & \cm        \\ \bottomrule
    \end{tabular}
    \caption{20 most common tracking domains after ATT: sharing of user identifiers with third-parties, alongside device information. Empty cells mean that we did not observe the sharing of a certain type of information, although this might still take place.}
    \label{tab:id_sharing}
\end{table*}

In our analysis, we found 9 apps that were able to generate a mutual user identifier that can be used for cross-app tracking, through the use of server-side code.
These 9 apps used an \enquote{AAID} (potentially leaning on the term Android Advertising Identifier) implemented and generated by Umeng, a subsidiary of the Chinese tech company Alibaba.
The flow to obtain an AAID is visualised in Figures~\ref{fig:aaid1} and \ref{fig:aaid2} in the Appendix.
As expected, the IDFA is only zeros because we used the opt-out provided by iOS 14.8; we observed, however, that the IDFV (ID for Vendors), a non-resettable, app-specific identifier was shared over the Internet, see Figure~\ref{fig:aaid1}.
The sharing of device information for purposes of fingerprinting would be in violation of the Apple's policies, which do not allow developers to \enquote{derive data from a device for the purpose of uniquely identifying it}~\cite{apple_tracking_definition}.
Other experts and researchers have also voiced concerns that tracking might continue~\cite{att_caid1,apple_enforcement1,apple_enforcement2,apple_enforcement3}.

We reported our observations to Apple on 17 November 2021, who promised to investigate the problem.
We conducted a follow-up investigation on 1 February 2022, and re-downloaded and analysed a range of iOS apps.
Some of the apps still continued to retrieve a unique identifier from the URL \url{https://aaid.umeng.com/api/postZdata}.
Other apps now contacted the URL \url{https://utoken.umeng.com/api/postZdata/v2}, and applied additional encryption (rather than just HTTPS) to the requests and responses.
This encrypted data had roughly the same size as before (\textasciitilde750 bytes for the request, \textasciitilde350 bytes for the response) and the same mimetype (\path{application/json} for the request, \path{application/json;charset=UTF-8} for the response).
The issue seems thus to be present still, but has now been hidden away from the public through the use of encryption.
We have tried to reproduce these experiments for a few apps on iOS 15 and higher, but did not observe the same behaviour; there currently exists no public jailbreak for these iOS versions, and similar investigations as ours are therefore not (yet) possible on these iOS versions.
There is a possibility that the issue has been fixed on iOS 15 or higher, or that we did not pick up the same behaviour in our small-scale testing (about 10 apps instead of more than 1000).
However, Apple did not provide further details to us.

Analysing the top 20 most commonly contacted domains, we could confirm that installation-specific identifiers (see column \enquote{User ID}) are commonly collected alongside further device-specific information, see Table~\ref{tab:id_sharing}.
While these installation-specific identifiers are usually randomly generated at the first app start, large tracking companies can likely still use these identifiers to build profiles of an app user's journey across apps, using their server-side code to link different identifiers together (e.g. through the user's IP address, other device information, and first-party data).
Companies also receive information about a user's locale (i.e. the display language), the device model, and the OS version.
Such information can be used to disambiguate different users connecting from the same IP address (e.g. households sharing the same Wi-Fi router)~--~and even across different IP addresses through the use of additional, first-party data that large tracking companies hold.

Table~\ref{tab:id_sharing} does not include all the different kinds of information that we observed being sent to tracking domains because the kinds of information varied between companies.
For example, Google assigned an \texttt{android\_id} to an iOS app upon first contact with the company that was then used for all subsequent communication with Google's API endpoints. This identifier differed between apps, and did not seem to be used for cross-app tracking on-device (it might be on Google's servers).
When contacting the domain \path{googleads.g.doubleclick.net}, Google collected the current system volume and the status of the silencing button.
As already described above, \path{ca.iadsdk.apple.com} collected a \texttt{purchaseTimestamp}, that can be used to identify the user, and is not accessible for other app developers.
The domain \texttt{gsp64-ssl.ls.apple.com}, belonging to Apple's location services, even collected the IP address and port that we used for proxying the network traffic through \texttt{mitmdump} as part of our analysis.
We did not observe any other domains that had access to this information, underlining Apple's privileged data access.
Crucially, for many of the observed transmissions between apps and servers, we could not even determine what data was sent, due to use of encryption~\cite{apple_enforcement2} and closed-source communication protocols.

\textbf{System-Level Tracking by Apple.}
We found that iPhones exchanged a range of unique user identifiers directly with Apple, see Figure~\ref{fig:apple_id_collection} in the Appendix.
We observed that network requests, which included various unique user identifiers and other personal data, were issued following the interaction with apps and connected to Apple's App Store and advertising technologies.
While this does not allow user-level apps to gain access to these user identifiers, Apple itself can use these identifiers to enrich its own advertising services.
Indeed, Apple claims in its privacy policy that it may use users' interactions with its advertising platform and with the App Store to group users into segments (of at least 5,000 individuals), and show adverts to these groups~\cite{apple_advertising_tracking}.
Specifically, we found that the App Store collected the UDID, the serial number of the device, the DSID (an identifier linked to a user's Apple account), and a \texttt{purchaseTimestamp}.
All of these identifiers can be used by Apple to single out individual users.
Crucially, the UDID has been inaccessible to app developers other than Apple since 2013~\cite{uuid_deprecation}, but Apple continues to have access to this identifier.
Moreover, Apple collects the serial number, which cannot be changed and is linked to a user's iPhone.
This might be unexpected for some users.
These findings are in-line with previous reports that both
Google and Apple collect detailed information about their users as part of regular device usage~\cite{leith_mobile_2021}.

\subsection{Disclosure of Tracking in Privacy Nutrition Labels}
\label{sec:nutrition_labels}

We now consider whether and to what extent apps (from after the introduction of iOS 14) disclose their tracking activities in their Privacy Nutrition Labels.

\begin{figure}
    \centering
	\includegraphics[width=0.95\linewidth]{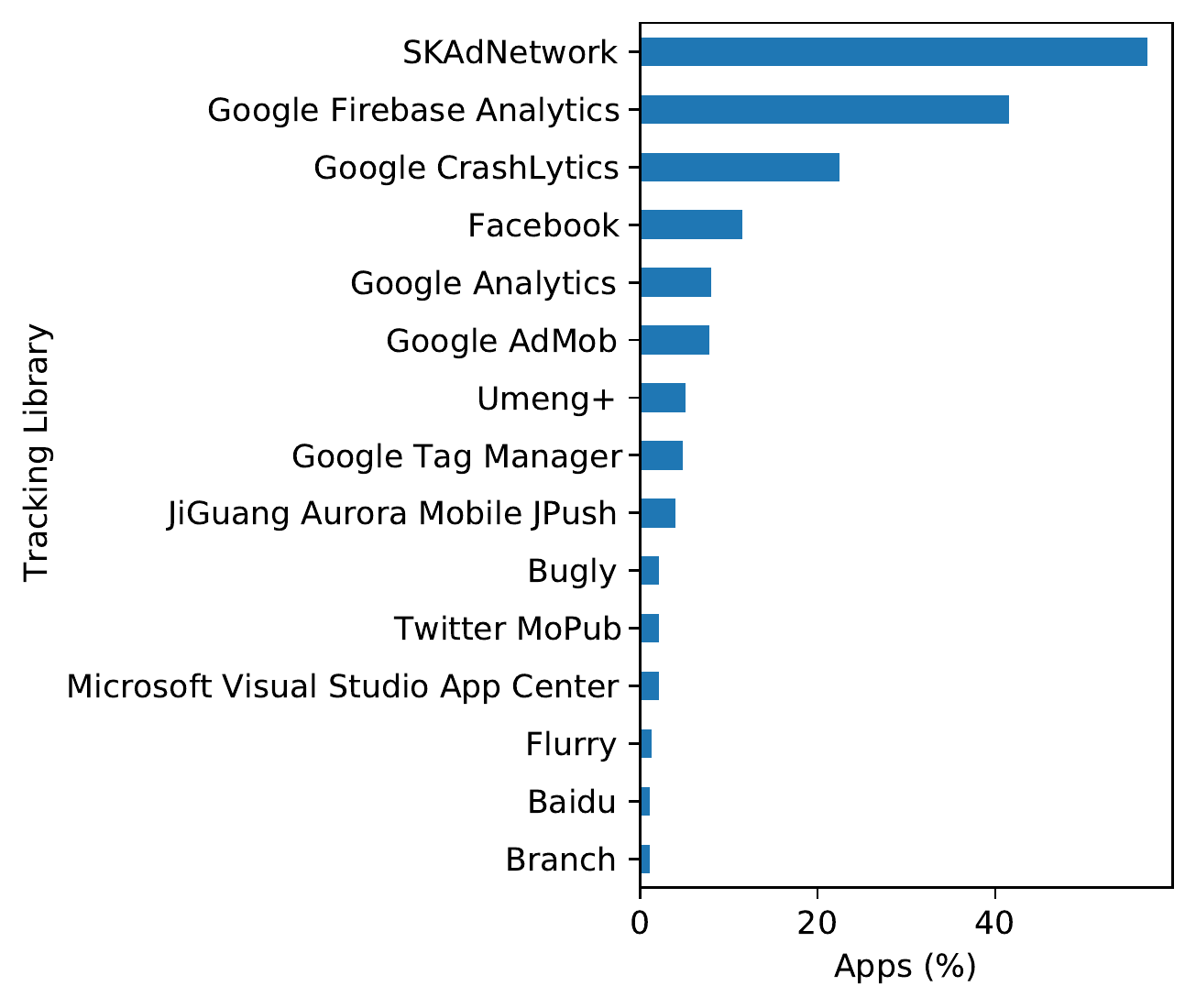}
    \caption{Top tracking libraries in apps that claim in their Privacy Nutrition Labels not to collect any data.}
    \label{fig:datanotcollected}
    \Description{This figure shows a bar chart of the top tracking libraries in analysed apps that claim in their Privacy Nutrition Labels not to collect any data.}
\end{figure}

Among the studied apps, 22.2\% claimed that they would not collect any data from the user.
This was often not true: as shown in Figure~\ref{fig:datanotcollected},
80.2\% of these apps actually contained at least one tracker library (compared to 93.1\% for apps that did disclose some data sharing), and 68.6\% sent data to at least one known tracking domain right at the first app start (compared to 91.4\%).
On average, apps that claimed not to collect data contained 1.8 tracking libraries (compared to 4.3), and contacted 2.5 tracking companies (compared to 4.2).
Among the 22.2\% of apps claiming not to collect data, only 3 were in the App Store charts. 
As noticed above (see Table~\ref{tab:id_sharing}), tracking libraries usually create a unique user identifier.
Among the apps that used the SKAdNetwork, 42.0\% disclosed their access to a \enquote{User ID}, 42.2\% of apps using Google Firebase Analytics, 48.2\% of apps using Google Crashlytics, and 53.2\% of apps using the Facebook SDK.
63.2\% of apps using Google Firebase Analytics disclosed that they collected any data about \enquote{Product Interaction} or \enquote{Other Usage Data}, and about 70\% of apps using the Facebook SDK, Google Analytics, or Google Tag Manager.
Additionally, apps can disclose their use of \enquote{Advertising Data}: 27.5\% of apps with the SKAdNetwork did so, 66.0\% of apps with Google AdMob, 80.9\% of apps with Unity3d Ads, and 45.4\% apps with AppsFlyer.

All of this points to notable discrepancies between apps' disclosed and actual data practices.
App developers might be able to address this, but are often not fully aware of all the data that is collected through third-party tracking software~\cite{mhaidli_we_2019,anirudhchi2021}.
Conversely, Apple itself might be able to reduce this discrepancy through increased use of automated code analysis, in particular applied to third-party tracking software.

%% file: 5_discussion.tex
\textbf{Tracking continues, and reinforces the power of gatekeepers and opacity of the mobile data ecosystem.}
Our findings suggest that tracking companies, especially larger ones with access to large troves of first-party data, can still track users behind the scenes.
They can do this through a range of methods, including using IP addresses to link installation-specific IDs across apps and through the sign-in functionality provided by individual apps (e.g. Google or Facebook sign-in, or email address).
Especially in combination with further user and device characteristics, which our data confirmed are still widely collected by tracking companies, it would be possible to analyse user behaviour across apps and websites (i.e. fingerprinting and cohort tracking).
A direct result of the ATT could therefore be that existing power imbalances in the digital tracking ecosystem get reinforced.

We even found a real-world example of Umeng, a subsidiary of the Chinese tech company Alibaba, using their server-side code to provide apps with a fingerprinting-derived cross-app identifier, see Figure~\ref{fig:fingerprinting} in the Appendix.
The use of fingerprinting is in violation of Apple’s policies~\cite{apple_tracking_definition}, and raises questions around the extent to which Apple can enforce its policies against server-side code.
ATT might ultimately encourage a shift of tracking technologies behind the scenes, so that they are outside of Apple's reach.
In other words, Apple's new rules might lead to even less transparency around tracking than we currently have, including for academic researchers.

\textbf{Privacy Nutrition Labels can be inaccurate and misleading, and have so far not changed data practices.}
Our results suggest that there is a discrepancy between apps' disclosed (in their Privacy Nutrition Labels) and actual data practices.
We observed that many (mostly less popular) apps gave incomplete information or falsely declared not to collect any data at all.
These observations are not necessarily to blame on app developers, who often have no idea of how third-party libraries handle users' personal data~\cite{kollnig_2021,anirudhchi2021,mhaidli_we_2019}.
As reported in Section~\ref{sec:static_tracking_config}, the proportion of app developers that make use of data-minimising settings of popular tracker libraries has roughly doubled, but these developers still remain a small minority.
The Privacy Nutrition Labels have not (yet) had an impact on developers' actual practices at large, but might do so in the long run by both increasing app users' privacy expectations and making app developers rethink their privacy practices~\cite{kelley_nutrition_2009,10.1145/1753326.1753561}.
As they stand, the labels can be misleading and create a false sense of security for consumers.

\textbf{Are the most egregious and opaque trackers tamed now?}
The reduced access to permanent user identifiers through ATT
could substantially improve app privacy.
While in the short run, some companies might try to replace the IDFA with statistical identifiers,
the reduced access to non-probabilistic cross-app identifiers might make it very hard for data brokers and other smaller tracker companies to compete.
Techniques like fingerprinting and cohort tracking may end up not being competitive enough compared to more privacy-preserving, on-device solutions.
We are already seeing a shift of the advertising industry towards the adoption of such solutions, driven by decisions of platform gatekeepers (e.g. Google's FloC / Topics API and Android Privacy Sandbox, Apple's ATT and Privacy Nutrition Labels)~\cite{kollnig_before_2021,att_cohorts}, though more discussion is needed around the effectiveness of these privacy-protecting  technologies.
The net result, however, of this shift towards more privacy-preserving methods is likely going to be more concentration with the existing platform gatekeepers, as the early reports on the tripled marketing share of Apple~~\cite{att_impact_revenues}, the planned overhaul of advertising technologies by Facebook/Meta and others~\cite{att_cohorts}, and the shifting spending patterns of advertisers suggest~\cite{att_marketingbudgets}.
Advertising to iOS users~--~being some of the wealthiest individuals~--~will be an opportunity that many advertisers cannot miss out on, and so they will rely on the advertising technologies of the larger tech companies to continue targeting the right audiences with their ads.

\textbf{Failure of GDPR enforcement, and power of platforms.}
Apple's new rules should not have a dramatic effect on the tracking of users in the EU and UK, given that existing data protection laws in these jurisdictions already ban most forms of third-party tracking without user consent~\cite{kollnig_2021,apps_consent_2021}.
While there was vocal outcry over Apple's new privacy measures by advertisers, the adtech industry was aware of tightened EU and UK data protection rules since April 2016, and had plenty of time to work out a way to ensure compliance with basic provisions of the GDPR, until May 2018, including the need to seek consent from users before engaging in tracking~\cite{kollnig_2021}.
Broad empirical evidence, from this and other pieces of research~~\cite{kollnig2021iphones,kollnig_tracking_2019,kollnig_2021,okoyomon_ridiculousness_2019,maps_2019,reyes_wont_2018}, suggests that apps' compliance with the GDPR is somewhat limited.

At the same time, it is worrying that a few changes by a private company (Apple) seem to have changed data protection in apps more than many years of high-level discussion and efforts by regulators, policymakers and others. This highlights the relative power of these gatekeeper companies, and the failure of regulators thus far to enforce the GDPR adequately.
An effective approach to increase compliance with data protection law and privacy protections in practice might be more targeted regulation of the gatekeepers of the app ecosystem; so far, there exists no targeted regulation in the US, UK and EU (see Section~\ref{sec:regulation}).

\textbf{Apple's Double Standards I: Making and Enforcing App Store Policies.}
Our analysis shows that Apple has a competitive advantage within the iOS ecosystem in various ways.
First, it both makes the rules for the App Store and interprets them in practice.
This is reflected in Apple's definition of tracking, which ostensibly exempts its own advertising technology~\cite{apple_advertising_tracking}:
\enquote{Tracking refers to the act of linking user or device \textit{data collected from your app} with user or device data collected from \textit{other companies’ apps, websites, or offline properties} for \textit{targeted advertising or advertising measurement purposes}. Tracking also refers to sharing user or device data with \textit{data brokers}.} (emphasis added)~\cite{apple_tracking_definition}
In other words, for tracking to fall under Apple's definition, it must fulfil three conditions, or be done by a data broker.

Apple's definition hinges on a distinction between first-party and third-party data collection, when this is not usually the root of privacy problems.
This is why the W3C defines tracking as \enquote{the collection of data regarding a particular user's activity across multiple distinct contexts and the retention, use, or sharing of data derived from that activity outside the context in which it occurred.}~\cite{w3c_tracking_definition}.
Rather than \textit{companies}, this definition is centred around different \textit{contexts}, as is commonly sought to be protected in privacy theory (e.g. contextual integrity~\cite{nissenbaum_privacy_2004}) and in privacy and data protection law (e.g. purpose limitation under Article 5 of the GDPR).
Apple's definition of tracking might both betray the expectation of consumers who expect that tracking would stop (when first-party tracking, notably by Apple itself, continues to be allowed), and motivate other companies to consolidate and join forces leading to increased market concentration.

Apple additionally foresees a list of exempt practices~\cite{apple_tracking_definition} (see Figure~\ref{fig:exemptions} in the Appendix for an excerpt).
These include \enquote{fraud detection, fraud prevention, or security purposes}, which might be interpreted extremely broadly by tracking companies.
The exempt practices further allow tracking by a \enquote{consumer reporting agency}.
The term \enquote{consumer reporting agency} is defined in the US Fair Credit Reporting Act (FCRA), regulating the relationship between these agencies and other \enquote{furnishers of information} relating to consumers.
By explicitly exempting credit scoring, Apple might try to avoid liability, and it might not have much choice under current US law.
The exemption of credit scoring is nonetheless problematic because the use of personal data for credit scoring can have disproportionate impacts on individuals, and might be protected by other data protection and privacy laws.
This might create the (false) impression for some app developers that other legal conditions do not apply, and a \textit{false sense of security} for many consumers.

\textbf{Apple's Double Standards II: Access to Data.}
Being the maker of the iOS ecosystem, Apple has a certain competitive advantage, by being able to collect device and user data, including hardware identifiers, that other app developers do not have access to, and use this for its own business purposes.
For example, by collecting the device's serial number regularly, Apple can accurately tie the point-of-sale of its devices to activities on the device itself, and track the device lifecycle in great detail.
Some of Apple's own apps, including the App Store itself, have access to this information because they are not distributed via the App Store and hence do not fall under the rules governing the App Store, including those that relate to tracking of users.
These observations support the known concerns around fair competition in the App Store. 

\subsection{Limitations}
\label{sec:limitations}

A few limitations of our study are worth noting. First, for practical reasons, we were not able to analyse all the apps in the App Store, only a reasonably large subset of free apps in the App Store's UK region.
Furthermore, for the purposes of examining the effect of ATT,
we only focused on apps that already existed on the App Store before iOS 14~--~newly released apps may adopt different strategies.
Regarding our analysis methods, our instruments are also potentially limited in several ways.  The results of our static analysis must be interpreted with care, since not all code shipped in an app will necessarily be invoked in practice. We may have overestimated tracking in certain contexts, e.g., if tracking code was included but not used.
In our network analysis, we performed this off-device, meaning that all device traffic was analysed in aggregate.  The risk here is that we may wrongly attribute some communications to an app that in fact was generated by some other app or subsystem on the device. To  minimise this risk, we uninstalled all pre-installed apps, and ensured no apps were running in the background. 
We also used jailbreaking (i.e. gained full system access by exploiting a vulnerability in the iOS operating system) to circumvent certificate validation, which might make some apps alter their behaviour.
In all parts of our analysis, we consider all apps equally,
regardless of popularity~\cite{binns_measuring_2018} and usage time~\cite{van_kleek_x-ray_2018}, both of which can impact user privacy.
Likewise, we treat all tracking domains, libraries and companies equally, though they might pose different risks to users.

%% file: 6_conclusions.tex
Overall, we find that Apple's new policies largely live up to its promises on making tracking more difficult.
Tracking libraries cannot access the IDFA anymore, and this directly impacts the business of data brokers.
These data brokers can pose significant risks to individuals, since they try to amass data about individuals from a wide range of contexts and sell this information to third-parties.
At the same time, apps still widely use tracking technology of large companies, and send a range of user and device characteristics over the Internet for the purposes of cohort tracking and user fingerprinting.
We found real-world evidence of apps computing a mutual fingerprinting-derived identifier through the use of server-side code (see Section~\ref{sec:data_sharing_pii} and Figure~\ref{fig:fingerprinting} in the Appendix)~--~a violation of Apple's new policies~\cite{apple_tracking_definition}, highlighting limits of Apple's enforcement power as a privately-owned data protection regulator~\cite{hoboken2021,greene_platform_2018}.
Indeed, Apple itself engages in some forms of user tracking (see Section~\ref{sec:data_sharing_pii} and Figure~\ref{fig:apple_id_collection}) and exempts invasive data practices like first-party tracking and credit scoring from its definition of tracking.
Lastly, we found the Privacy Nutrition Labels to be sometimes incomplete and inaccurate, especially in less popular apps (Section \ref{sec:nutrition_labels}).

Apple's privacy changes have led to positive improvements for user privacy.
However, we also found various aspects that are in conflict with Apple's marketing claims and might go against users' reasonable privacy expectations, e.g. that the new opt-in tracking prompts would stop all tracking, that the new Privacy Nutrition Labels would always be correct and be verified by Apple, or that Apple would be subject to the same restrictions to data access and privacy rules as other companies.
There is a risk that individuals will develop even more resignation over the use of their data online if they are provided with with misleading or ineffective privacy solutions~\cite{shklovski_leakiness_2014,colnago_informing_2020}.
This resignation could in the long run undermine privacy efforts and adversely affect fundamental rights, such as the rights to data protection and privacy. 

Despite positive developments over the recent months and years, especially through initiatives by Apple, there is still some way to go for app privacy.
Violations of various aspects of data protection and privacy laws remain widespread in apps~\cite{kollnig2021iphones,kollnig_tracking_2019,kollnig_2021,okoyomon_ridiculousness_2019,maps_2019,reyes_wont_2018}, while enforcement of existing data protection laws against such practices stays sporadic.
Apple's privacy efforts are hampered by its closed-source philosophy on iOS and the opacity around the enforcement of its App Store review policies.
To strengthen iOS privacy, Apple has already started to prevent IP-based tracking by routing traffic to trackers via its own servers when using the iOS browser (\enquote{Privacy Relay}).
As a direct response to our findings, Apple could consider extending the Privacy Relay to tracking within apps, thereby making the tracking of users through their IP address more difficult~\cite{apple_enforcement1}.
However, this would also further extend Apple's reach over the iOS ecosystem and potentially allow the company to track users even more accurately.

More generally, the key decision makers with regards to privacy technologies must establish robust transparency and accountability measures that allow for independent assessment of any privacy guarantees and promises.
This is especially true, given the current lack of targeted regulations for app platforms like Google Play and the Apple App Store (see Section~\ref{sec:regulation}).
In the case of Apple, improved transparency measures must necessarily involve the phasing out of encryption of free iOS apps by default, which currently forces independent privacy researchers into legal grey areas and severely hampers such research efforts (see Section~\ref{sec:related-work}).
This is why most previous privacy research focused on Android and the last large-scale privacy study into iOS apps had been conducted in 2013~\cite{agarwal_protectmyprivacy_2013}, until the recent release of the method used in this study~\cite{kollnig2021iphones}.

We conclude that the new changes by Apple have traded more privacy for more concentration of data collection with fewer tech companies.
Stricter privacy rules may encourage even less transparency around app tracking, by shifting tracking code onto the servers of dominant tracking companies.
Despite the new rules, large companies, like Google/Alphabet and Facebook/Meta, are still able to track users across apps, because these companies have access to unique amounts of first-party data about users.
Apple is now able to track its customers even more accurately, by taking a larger share in advertising technologies and getting unique access to user identifiers, including the device serial number.
This underlines that privacy and competition problems can be highly intertwined in digital markets and need holistic study.

\textbf{Future work.}
In this work, we only analysed apps that were already present on the App Store before iOS 14 and the ATT; it would be interesting to analyse how the ATT has impacted the privacy properties of \textit{newly released} apps on the App Store. 
It would also be helpful to develop a new automation tool for iOS apps to observe apps' data practices automatically, even beyond the first app start~--~as studied in this paper. 
It would be pertinent to study user tracking by platforms in more detail, and also how the Privacy Nutrition Labels inform individuals around app privacy.

%% file: 7_appendix.tex
\section*{Appendix}

\begin{figure}[H]
    \centering
    \includegraphics[width=0.7\linewidth]{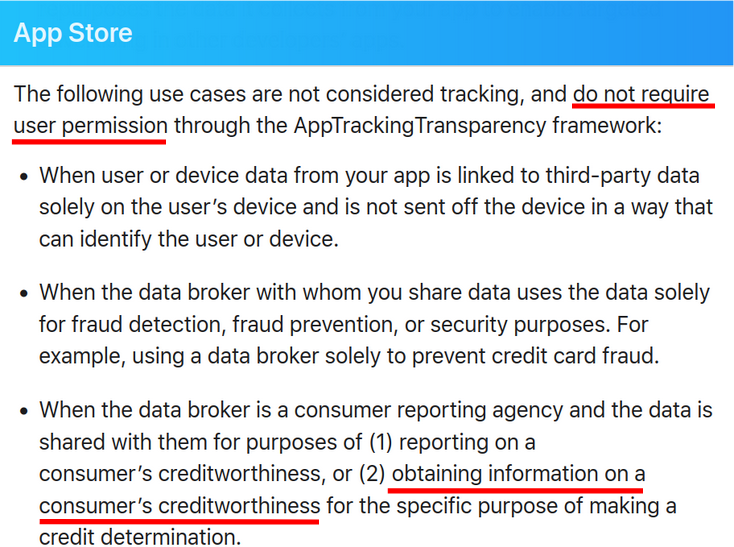}
    \caption{Apple's definition of tracking: Excerpt from Apple's exempt data practices, including credit scoring, from requiring user opt-in under ATT (emphasis added)~\cite{apple_tracking_definition}. We discuss the limitations of Apple's definition of tracking in Section~\ref{sec:discussion}.}
    \Description{This figure shows an excerpt from Apple's ATT rules that foresee an exemption for data collection related to credit scoring.}
    \label{fig:exemptions}
\end{figure}

\begin{figure}[H]
    \begin{subfigure}[t]{\linewidth}
        \lstset{language=json}
        \begin{lstlisting}
{
    "sdk_version": "1.2.0",
    "bundle_id": "[Redacted]",
    "hw_model": "N69uAP",
    "kid": "[Redacted]",
    "total_storage": "30745123781",
    "country": "GB",
    "zdata": "[Redacted]",
    "app_version": "[Redacted]",
    "app_name": "[Redacted]",
    "sdk_type": "IOS",
    "storage": "14078912372",
    "zdata_ver": "1.1.0",
    "source_id": "umeng",
    "idfv": "7EBDAFC8-97BB-4FDB-B4D3-E2F4EA040B8C",
    "timezone": "1",
    "os_version": "14.8",
    "model": "iPhone8,4",
    "hostname": "MyPhone",
    "appkey": "[Redacted]",
    "idfa": "00000000-0000-0000-0000-
        000000000000"
}
        \end{lstlisting}
        \caption{Request: Sending a range of device information to Umeng at \url{https://aaid.umeng.com/api/postZdata}.}
        \label{fig:aaid1}
	\end{subfigure}
	\begin{subfigure}[t]{\linewidth}
        \centering
        \lstset{language=json}
        \begin{lstlisting}
{
    "aaid": "BAEC362C-49FC-494B-B0A7-175D990B059D",
    ...
}
        \end{lstlisting}
        \caption{Response: Umeng returns an identifier that is shared by multiple apps, and can be used for cross-app tracking.}
        \label{fig:aaid2}
    \end{subfigure}
	\caption{Fingerprinting in apps, even after the ATT. This is likely in violation of Apple's new policies and the expectations of many end-users (personal data redacted). We provide more results on the circumvention of the ATT in Section~\ref{sec:data_sharing_pii}.}\label{fig:fingerprinting}
	\Description{This figures shows the content of a network request and of the subsequent response of a tracking company helping apps agree on a shared unique use identifier.}
\end{figure}

\begin{figure}[H]
	\begin{subfigure}[t]{\linewidth}
        \centering
        \lstset{language=XML}
        \begin{lstlisting}
<plist version="1.0">
<dict>
	...
	<key>dsid</key>
	<string>[Apple ID]</string>
	<key>guid</key>
	<string>[UDID]</string>
	<key>serialNumber</key>
	<string>[serial number]</string>
	...
</dict>
</plist>
        \end{lstlisting}
        \caption{Request of Apple App Store to \url{https://buy.itunes.apple.com/WebObjects/MZFinance.woa/wa/renewVppReceipt?guid=[UDID]}.}
	\end{subfigure}
	\begin{subfigure}[t]{\linewidth}
        \centering
        \lstset{language=json}
        \begin{lstlisting}
{
    "attributionMetadataExistsOnDevice": false,
    "toroId": "[Redacted]",
    "purchaseTimestamp": "2021-11-01T15:15:05Z",
    "adamId": 477718890,
    "attributionDownloadType": 0,
    "developmentApp": false,
    "anonymousDemandId": "[Redacted]",
    "bundleId": "ru.kinopoisk",
    "attributionKey": "[Redacted]"
}
        \end{lstlisting}
        \caption{Request (shortended) of Apple's advertising framework to \url{https://ca.iadsdk.apple.com/adserver/attribution/v2}.}
    \end{subfigure}
	\caption{Sharing of unique user identifiers with Apple (personal data redacted). We explain more about the tracking of users by Apple in Section~\ref{sec:data_sharing_pii}.}~\label{fig:apple_id_collection}
	\Description{This figures shows two example network traffic of Apple collecting personal data that can be used for user tracking.}
\end{figure}

%% file: 0_main.bbl

\begin{thebibliography}{63}


\ifx \showCODEN    \undefined \def \showCODEN     #1{\unskip}     \fi
\ifx \showDOI      \undefined \def \showDOI       #1{#1}\fi
\ifx \showISBNx    \undefined \def \showISBNx     #1{\unskip}     \fi
\ifx \showISBNxiii \undefined \def \showISBNxiii  #1{\unskip}     \fi
\ifx \showISSN     \undefined \def \showISSN      #1{\unskip}     \fi
\ifx \showLCCN     \undefined \def \showLCCN      #1{\unskip}     \fi
\ifx \shownote     \undefined \def \shownote      #1{#1}          \fi
\ifx \showarticletitle \undefined \def \showarticletitle #1{#1}   \fi
\ifx \showURL      \undefined \def \showURL       {\relax}        \fi
\providecommand\bibfield[2]{#2}
\providecommand\bibinfo[2]{#2}
\providecommand\natexlab[1]{#1}
\providecommand\showeprint[2][]{arXiv:#2}

\bibitem[\protect\citeauthoryear{Agarwal and Hall}{Agarwal and Hall}{2013}]%
        {agarwal_protectmyprivacy_2013}
\bibfield{author}{\bibinfo{person}{Yuvraj Agarwal} {and}
  \bibinfo{person}{Malcolm Hall}.} \bibinfo{year}{2013}\natexlab{}.
\newblock \showarticletitle{{{ProtectMyPrivacy}}: Detecting and Mitigating
  Privacy Leaks on {{iOS}} Devices Using Crowdsourcing}. In
  \bibinfo{booktitle}{\emph{Proceeding of the 11th Annual International
  Conference on {{Mobile}} Systems, Applications, and Services - {{MobiSys}}
  '13}}. \bibinfo{publisher}{{ACM Press}}, \bibinfo{address}{{Taipei, Taiwan}},
  \bibinfo{pages}{97}.
\newblock
\showISBNx{978-1-4503-1672-9}
\urldef\tempurl%
\url{https://doi.org/10.1145/2462456.2464460}
\showDOI{\tempurl}


\bibitem[\protect\citeauthoryear{{Apple}}{{Apple}}{2021a}]%
        {apple_advertising_tracking}
\bibfield{author}{\bibinfo{person}{{Apple}}.} \bibinfo{year}{2021}\natexlab{a}.
\newblock \bibinfo{title}{{Apple Advertising \& Privacy}}.
\newblock
  \bibinfo{howpublished}{\url{https://www.apple.com/legal/privacy/data/en/apple-advertising/}}.
\newblock


\bibitem[\protect\citeauthoryear{{Apple}}{{Apple}}{2021b}]%
        {apple_tracking_definition}
\bibfield{author}{\bibinfo{person}{{Apple}}.} \bibinfo{year}{2021}\natexlab{b}.
\newblock \bibinfo{title}{{User Privacy and Data Use}}.
\newblock
  \bibinfo{howpublished}{\url{https://developer.apple.com/app-store/user-privacy-and-data-use/}}.
\newblock


\bibitem[\protect\citeauthoryear{{AppsFlyer}}{{AppsFlyer}}{2021}]%
        {att_optout3}
\bibfield{author}{\bibinfo{person}{{AppsFlyer}}.}
  \bibinfo{year}{2021}\natexlab{}.
\newblock \bibinfo{title}{{Initial data indicates ATT opt-in rates are much
  higher than anticipated — at least 41\%}}.
\newblock
  \bibinfo{howpublished}{\url{https://www.appsflyer.com/blog/trends-insights/att-opt-in-rates-higher/}}.
\newblock


\bibitem[\protect\citeauthoryear{{Authority for Consumers and
  Markets}}{{Authority for Consumers and Markets}}{2022}]%
        {dutch_authority}
\bibfield{author}{\bibinfo{person}{{Authority for Consumers and Markets}}.}
  \bibinfo{year}{2022}\natexlab{}.
\newblock \bibinfo{title}{{ACM obliges Apple to adjust unreasonable conditions
  for its App Store}}.
\newblock
  \bibinfo{howpublished}{\url{https://www.acm.nl/en/publications/acm-obliges-apple-adjust-unreasonable-conditions-its-app-store}}.
\newblock


\bibitem[\protect\citeauthoryear{Binns, Lyngs, Van~Kleek, Zhao, Libert, and
  Shadbolt}{Binns et~al\mbox{.}}{2018a}]%
        {binns_third_2018}
\bibfield{author}{\bibinfo{person}{Reuben Binns}, \bibinfo{person}{Ulrik
  Lyngs}, \bibinfo{person}{Max Van~Kleek}, \bibinfo{person}{Jun Zhao},
  \bibinfo{person}{Timothy Libert}, {and} \bibinfo{person}{Nigel Shadbolt}.}
  \bibinfo{year}{2018}\natexlab{a}.
\newblock \showarticletitle{Third Party Tracking in the Mobile Ecosystem}. In
  \bibinfo{booktitle}{\emph{Proceedings of the 10th {ACM} Conference on Web
  Science - {WebSci} '18}} (Amsterdam, Netherlands). \bibinfo{publisher}{{ACM}
  Press}, \bibinfo{address}{New York, NY, USA}, \bibinfo{pages}{23--31}.
\newblock
\showISBNx{978-1-4503-5563-6}
\urldef\tempurl%
\url{https://doi.org/10.1145/3201064.3201089}
\showDOI{\tempurl}


\bibitem[\protect\citeauthoryear{Binns, Zhao, Kleek, and Shadbolt}{Binns
  et~al\mbox{.}}{2018b}]%
        {binns_measuring_2018}
\bibfield{author}{\bibinfo{person}{Reuben Binns}, \bibinfo{person}{Jun Zhao},
  \bibinfo{person}{Max~Van Kleek}, {and} \bibinfo{person}{Nigel Shadbolt}.}
  \bibinfo{year}{2018}\natexlab{b}.
\newblock \showarticletitle{Measuring Third-party Tracker Power across Web and
  Mobile}.
\newblock \bibinfo{journal}{\emph{{ACM} Transactions on Internet Technology}}
  \bibinfo{volume}{18}, \bibinfo{number}{4} (\bibinfo{year}{2018}),
  \bibinfo{pages}{1--22}.
\newblock
\showISSN{15335399}
\urldef\tempurl%
\url{https://doi.org/10.1145/3176246}
\showDOI{\tempurl}


\bibitem[\protect\citeauthoryear{Bygrave}{Bygrave}{2017}]%
        {bygrave_data_2017}
\bibfield{author}{\bibinfo{person}{Lee~A Bygrave}.}
  \bibinfo{year}{2017}\natexlab{}.
\newblock \showarticletitle{Data Protection by Design and by Default:
  Deciphering the {EU}’s Legislative Requirements}.
\newblock \bibinfo{journal}{\emph{Oslo Law Review}}  \bibinfo{volume}{1}
  (\bibinfo{year}{2017}), \bibinfo{pages}{105--120}.
\newblock
\urldef\tempurl%
\url{https://doi.org/10.18261/issn.2387-3299-2017-02-03}
\showDOI{\tempurl}


\bibitem[\protect\citeauthoryear{Chen, Wang, Chen, Wang, Lee, Wang, Ma, Wang,
  Zhang, and Zou}{Chen et~al\mbox{.}}{2016}]%
        {chen_following_2016}
\bibfield{author}{\bibinfo{person}{Kai Chen}, \bibinfo{person}{Xueqiang Wang},
  \bibinfo{person}{Yi Chen}, \bibinfo{person}{Peng Wang},
  \bibinfo{person}{Yeonjoon Lee}, \bibinfo{person}{XiaoFeng Wang},
  \bibinfo{person}{Bin Ma}, \bibinfo{person}{Aohui Wang},
  \bibinfo{person}{Yingjun Zhang}, {and} \bibinfo{person}{Wei Zou}.}
  \bibinfo{year}{2016}\natexlab{}.
\newblock \showarticletitle{Following {{Devil}}'s {{Footprints}}:
  {{Cross}}-{{Platform Analysis}} of {{Potentially Harmful Libraries}} on
  {{Android}} and {{iOS}}}. In \bibinfo{booktitle}{\emph{2016 {{IEEE
  Symposium}} on {{Security}} and {{Privacy}} ({{SP}})}}.
  \bibinfo{publisher}{{IEEE}}, \bibinfo{address}{{San Jose, CA}},
  \bibinfo{pages}{357--376}.
\newblock
\urldef\tempurl%
\url{https://doi.org/10.1109/SP.2016.29}
\showDOI{\tempurl}


\bibitem[\protect\citeauthoryear{Colnago, Feng, Palanivel, Pearman, Ung,
  Acquisti, Cranor, and Sadeh}{Colnago et~al\mbox{.}}{2020}]%
        {colnago_informing_2020}
\bibfield{author}{\bibinfo{person}{Jessica Colnago}, \bibinfo{person}{Yuanyuan
  Feng}, \bibinfo{person}{Tharangini Palanivel}, \bibinfo{person}{Sarah
  Pearman}, \bibinfo{person}{Megan Ung}, \bibinfo{person}{Alessandro Acquisti},
  \bibinfo{person}{Lorrie~Faith Cranor}, {and} \bibinfo{person}{Norman Sadeh}.}
  \bibinfo{year}{2020}\natexlab{}.
\newblock \showarticletitle{Informing the {{Design}} of a {{Personalized
  Privacy Assistant}} for the {{Internet}} of {{Things}}}. In
  \bibinfo{booktitle}{\emph{Proceedings of the 2020 {{CHI Conference}} on
  {{Human Factors}} in {{Computing Systems}}}}. \bibinfo{publisher}{{ACM}},
  \bibinfo{address}{{Honolulu HI USA}}, \bibinfo{pages}{1--13}.
\newblock
\showISBNx{978-1-4503-6708-0}
\urldef\tempurl%
\url{https://doi.org/10.1145/3313831.3376389}
\showDOI{\tempurl}


\bibitem[\protect\citeauthoryear{{Datenschutzkonferenz}}{{Datenschutzkonferenz}}{2021}]%
        {datenschutzkonferenz_orientierungshilfe_2021}
\bibfield{author}{\bibinfo{person}{{Datenschutzkonferenz}}.}
  \bibinfo{year}{2021}\natexlab{}.
\newblock \bibinfo{title}{Orientierungshilfe Der {{Aufsichtsbehörden}} Für
  {{Anbieter}} von {{Telemedien}}}.
\newblock
\newblock


\bibitem[\protect\citeauthoryear{Egele, Kruegel, Kirda, and Vigna}{Egele
  et~al\mbox{.}}{2011}]%
        {pios_2011}
\bibfield{author}{\bibinfo{person}{Manuel Egele}, \bibinfo{person}{Christopher
  Kruegel}, \bibinfo{person}{Engin Kirda}, {and} \bibinfo{person}{Giovanni
  Vigna}.} \bibinfo{year}{2011}\natexlab{}.
\newblock \showarticletitle{{{PiOS}}: Detecting Privacy Leaks in {{iOS}}
  Applications}. In \bibinfo{booktitle}{\emph{Proceedings of the Network and
  Distributed System Security Symposium (NDSS) 2011}}. \bibinfo{publisher}{{The
  Internet Society}}, \bibinfo{address}{San Diego, California},
  \bibinfo{numpages}{15}~pages.
\newblock


\bibitem[\protect\citeauthoryear{Ekambaranathan, Zhao, and
  Van~Kleek}{Ekambaranathan et~al\mbox{.}}{2021}]%
        {anirudhchi2021}
\bibfield{author}{\bibinfo{person}{Anirudh Ekambaranathan},
  \bibinfo{person}{Jun Zhao}, {and} \bibinfo{person}{Max Van~Kleek}.}
  \bibinfo{year}{2021}\natexlab{}.
\newblock \showarticletitle{“{{Money}} makes the world go around”:
  Identifying Barriers to Better Privacy in Children’s Apps From
  Developers’ Perspectives}. In \bibinfo{booktitle}{\emph{Conference on Human
  Factors in Computing Systems (CHI ’21)}} ({Yokohama, Japan}, 2021).
  \bibinfo{publisher}{{ACM Press}}, \bibinfo{address}{NY, USA},
  \bibinfo{pages}{1--24}.
\newblock
\urldef\tempurl%
\url{https://doi.org/10.1145/3411764.3445599}
\showDOI{\tempurl}


\bibitem[\protect\citeauthoryear{Enck, Gilbert, Chun, Cox, Jung, McDaniel, and
  Sheth}{Enck et~al\mbox{.}}{2010}]%
        {enck_taintdroid_2010}
\bibfield{author}{\bibinfo{person}{William Enck}, \bibinfo{person}{Peter
  Gilbert}, \bibinfo{person}{Byung-Gon Chun}, \bibinfo{person}{Landon~P. Cox},
  \bibinfo{person}{Jaeyeon Jung}, \bibinfo{person}{Patrick McDaniel}, {and}
  \bibinfo{person}{Anmol~N. Sheth}.} \bibinfo{year}{2010}\natexlab{}.
\newblock \showarticletitle{{{TaintDroid}}: {{An Information}}-Flow {{Tracking
  System}} for {{Realtime Privacy Monitoring}} on {{Smartphones}}}. In
  \bibinfo{booktitle}{\emph{Proceedings of the 9th {{USENIX Conference}} on
  {{Operating Systems Design}} and {{Implementation}}}}
  \emph{(\bibinfo{series}{{{OSDI}}'10})}. \bibinfo{publisher}{{USENIX}
  Association}, \bibinfo{address}{Vancouver, BC}, \bibinfo{pages}{393--407}.
\newblock


\bibitem[\protect\citeauthoryear{{Federal Trade Commission}}{{Federal Trade
  Commission}}{2013}]%
        {ftc_app_stores}
\bibfield{author}{\bibinfo{person}{{Federal Trade Commission}}.}
  \bibinfo{year}{2013}\natexlab{}.
\newblock \bibinfo{title}{Mobile Privacy Disclosures--Building Trust Through
  Transparency}.
\newblock
  \bibinfo{howpublished}{\url{https://www.ftc.gov/sites/default/files/documents/reports/mobile-privacy-disclosures-building-trust-through-transparency-federal-trade-commission-staff-report/130201mobileprivacyreport.pdf}}.
\newblock


\bibitem[\protect\citeauthoryear{{Financial Times}}{{Financial Times}}{2021a}]%
        {att_impact_revenues}
\bibfield{author}{\bibinfo{person}{{Financial Times}}.}
  \bibinfo{year}{2021}\natexlab{a}.
\newblock \bibinfo{title}{{Alphabet and Microsoft smash estimates with \$110bn
  revenue haul}}.
\newblock
  \bibinfo{howpublished}{\url{https://www.ft.com/content/273aeecb-57a8-40f8-a2ba-8a21a635b289}}.
\newblock


\bibitem[\protect\citeauthoryear{{Financial Times}}{{Financial Times}}{2021b}]%
        {att_cohorts}
\bibfield{author}{\bibinfo{person}{{Financial Times}}.}
  \bibinfo{year}{2021}\natexlab{b}.
\newblock \bibinfo{title}{{Apple reaches quiet truce over iPhone privacy
  changes}}.
\newblock
  \bibinfo{howpublished}{\url{https://www.ft.com/content/69396795-f6e1-4624-95d8-121e4e5d7839}}.
\newblock


\bibitem[\protect\citeauthoryear{{Financial Times}}{{Financial Times}}{2021c}]%
        {att_marketingbudgets}
\bibfield{author}{\bibinfo{person}{{Financial Times}}.}
  \bibinfo{year}{2021}\natexlab{c}.
\newblock \bibinfo{title}{{Apple’s privacy changes create windfall for its
  own advertising business}}.
\newblock
  \bibinfo{howpublished}{\url{https://www.ft.com/content/074b881f-a931-4986-888e-2ac53e286b9d}}.
\newblock


\bibitem[\protect\citeauthoryear{{Financial Times}}{{Financial Times}}{2021d}]%
        {att_caid1}
\bibfield{author}{\bibinfo{person}{{Financial Times}}.}
  \bibinfo{year}{2021}\natexlab{d}.
\newblock \bibinfo{title}{{China’s tech giants test way around Apple’s new
  privacy rules}}.
\newblock
  \bibinfo{howpublished}{\url{https://www.ft.com/content/520ccdae-202f-45f9-a516-5cbe08361c34}}.
\newblock


\bibitem[\protect\citeauthoryear{{Financial Times}}{{Financial Times}}{2021e}]%
        {att_impact_revenues2}
\bibfield{author}{\bibinfo{person}{{Financial Times}}.}
  \bibinfo{year}{2021}\natexlab{e}.
\newblock \bibinfo{title}{{Snap, Facebook, Twitter and YouTube lose nearly
  \$10bn after iPhone privacy changes}}.
\newblock
  \bibinfo{howpublished}{\url{https://www.ft.com/content/4c19e387-ee1a-41d8-8dd2-bc6c302ee58e}}.
\newblock


\bibitem[\protect\citeauthoryear{{Flurry}}{{Flurry}}{2021}]%
        {att_optout2}
\bibfield{author}{\bibinfo{person}{{Flurry}}.} \bibinfo{year}{2021}\natexlab{}.
\newblock \bibinfo{title}{{iOS 14.5 Opt-in Rate - Daily Updates Since Launch}}.
\newblock
  \bibinfo{howpublished}{\url{https://www.flurry.com/blog/ios-14-5-opt-in-rate-att-restricted-app-tracking-transparency-worldwide-us-daily-latest-update/}}.
\newblock


\bibitem[\protect\citeauthoryear{{Frida}}{{Frida}}{[n.\,d.]}]%
        {frida}
\bibfield{author}{\bibinfo{person}{{Frida}}.}
  \bibinfo{year}{[n.\,d.]}\natexlab{}.
\newblock \bibinfo{title}{{Frida: A world-class dynamic instrumentation
  framework}}.
\newblock \bibinfo{howpublished}{\url{https://frida.re}}.
\newblock


\bibitem[\protect\citeauthoryear{Greene and Shilton}{Greene and
  Shilton}{2018}]%
        {greene_platform_2018}
\bibfield{author}{\bibinfo{person}{Daniel Greene} {and} \bibinfo{person}{Katie
  Shilton}.} \bibinfo{year}{2018}\natexlab{}.
\newblock \showarticletitle{Platform privacies: Governance, collaboration, and
  the different meanings of “privacy” in {iOS} and Android development}.
\newblock \bibinfo{journal}{\emph{New Media \&amp; Society}}
  \bibinfo{volume}{20}, \bibinfo{number}{4} (\bibinfo{year}{2018}),
  \bibinfo{pages}{1640--1657}.
\newblock
\showISSN{1461-4448, 1461-7315}
\urldef\tempurl%
\url{https://doi.org/10.1177/1461444817702397}
\showDOI{\tempurl}


\bibitem[\protect\citeauthoryear{Han, Reyes, Elazari, Reardon, Feal, Bamberger,
  Egelman, and Vallina-Rodriguez}{Han et~al\mbox{.}}{2019}]%
        {free_v_paid_2019}
\bibfield{author}{\bibinfo{person}{Catherine Han}, \bibinfo{person}{Irwin
  Reyes}, \bibinfo{person}{Amit Elazari}, \bibinfo{person}{Joel Reardon},
  \bibinfo{person}{Alvaro Feal}, \bibinfo{person}{Kenneth~A. Bamberger},
  \bibinfo{person}{Serge Egelman}, {and} \bibinfo{person}{Narseo
  Vallina-Rodriguez}.} \bibinfo{year}{2019}\natexlab{}.
\newblock \showarticletitle{Do You Get What You Pay For? Comparing The Privacy
  Behaviors of Free vs. Paid Apps.}. In \bibinfo{booktitle}{\emph{The Workshop
  on Technology and Consumer Protection (ConPro ’19)}}.
  \bibinfo{publisher}{Institute of Electrical and Electronics Engineers},
  \bibinfo{address}{NY, USA}, \bibinfo{numpages}{7}~pages.
\newblock


\bibitem[\protect\citeauthoryear{Han, Reyes, Álvaro Feal, Reardon, Wijesekera,
  Vallina-Rodriguez, Elazari, Bamberger, and Egelman}{Han
  et~al\mbox{.}}{2020}]%
        {han_price_2020}
\bibfield{author}{\bibinfo{person}{Catherine Han}, \bibinfo{person}{Irwin
  Reyes}, \bibinfo{person}{Álvaro Feal}, \bibinfo{person}{Joel Reardon},
  \bibinfo{person}{Primal Wijesekera}, \bibinfo{person}{Narseo
  Vallina-Rodriguez}, \bibinfo{person}{Amit Elazari},
  \bibinfo{person}{Kenneth~A. Bamberger}, {and} \bibinfo{person}{Serge
  Egelman}.} \bibinfo{year}{2020}\natexlab{}.
\newblock \showarticletitle{The Price is (Not) Right: Comparing Privacy in Free
  and Paid Apps}.
\newblock \bibinfo{journal}{\emph{Proceedings on Privacy Enhancing
  Technologies}} \bibinfo{volume}{2020}, \bibinfo{number}{3}
  (\bibinfo{year}{2020}), \bibinfo{pages}{222--242}.
\newblock
\urldef\tempurl%
\url{https://doi.org/10.2478/popets-2020-0050}
\showDOI{\tempurl}


\bibitem[\protect\citeauthoryear{Han, Yan, Gao, Zhou, and Deng}{Han
  et~al\mbox{.}}{2013}]%
        {han_comparing_2013}
\bibfield{author}{\bibinfo{person}{Jin Han}, \bibinfo{person}{Qiang Yan},
  \bibinfo{person}{Debin Gao}, \bibinfo{person}{Jianying Zhou}, {and}
  \bibinfo{person}{Robert~H Deng}.} \bibinfo{year}{2013}\natexlab{}.
\newblock \showarticletitle{Comparing {{Mobile Privacy Protection}} through
  {{Cross}}-{{Platform Applications}}}. In
  \bibinfo{booktitle}{\emph{Proceedings 2013 {{Network}} and {{Distributed
  System Security Symposium}}}} ({San Diego, CA}).
  \bibinfo{publisher}{{Internet Society}}, \bibinfo{pages}{16}.
\newblock


\bibitem[\protect\citeauthoryear{{International Association of Privacy
  Professionals}}{{International Association of Privacy Professionals}}{2021}]%
        {att_optout1}
\bibfield{author}{\bibinfo{person}{{International Association of Privacy
  Professionals}}.} \bibinfo{year}{2021}\natexlab{}.
\newblock \bibinfo{title}{{Apple's ATT rollout presents uncertain path for
  adtech}}.
\newblock
  \bibinfo{howpublished}{\url{https://iapp.org/news/a/apples-att-rollout-presents-uncertain-path-for-adtech/}}.
\newblock


\bibitem[\protect\citeauthoryear{Jasmontaite, Kamara, Zanfir-Fortuna, and
  Leucci}{Jasmontaite et~al\mbox{.}}{2018}]%
        {jasmontaite_data_2018}
\bibfield{author}{\bibinfo{person}{Lina Jasmontaite}, \bibinfo{person}{Irene
  Kamara}, \bibinfo{person}{Gabriela Zanfir-Fortuna}, {and} \bibinfo{person}{S
  Leucci}.} \bibinfo{year}{2018}\natexlab{}.
\newblock \showarticletitle{Data {{Protection}} by {{Design}} and by
  {{Default}}: {{Framing Guiding Principles}} into {{Legal Obligations}} in the
  {{GDPR}}}.
\newblock \bibinfo{journal}{\emph{European Data Protection Law Review}}
  \bibinfo{volume}{4} (\bibinfo{year}{2018}), \bibinfo{pages}{168--189}.
\newblock
\urldef\tempurl%
\url{https://doi.org/10.21552/edpl/2018/2/7}
\showDOI{\tempurl}


\bibitem[\protect\citeauthoryear{Kelley, Bresee, Cranor, and Reeder}{Kelley
  et~al\mbox{.}}{2009}]%
        {kelley_nutrition_2009}
\bibfield{author}{\bibinfo{person}{Patrick~Gage Kelley},
  \bibinfo{person}{Joanna Bresee}, \bibinfo{person}{Lorrie~Faith Cranor}, {and}
  \bibinfo{person}{Robert~W. Reeder}.} \bibinfo{year}{2009}\natexlab{}.
\newblock \showarticletitle{A "Nutrition Label" for Privacy}. In
  \bibinfo{booktitle}{\emph{Proceedings of the 5th {{Symposium}} on {{Usable
  Privacy}} and {{Security}} - {{SOUPS}} '09}} ({Mountain View, California},
  2009). \bibinfo{publisher}{{ACM Press}}, \bibinfo{pages}{1}.
\newblock
\showISBNx{978-1-60558-736-3}
\urldef\tempurl%
\url{https://doi.org/10.1145/1572532.1572538}
\showDOI{\tempurl}


\bibitem[\protect\citeauthoryear{Kelley, Cesca, Bresee, and Cranor}{Kelley
  et~al\mbox{.}}{2010}]%
        {10.1145/1753326.1753561}
\bibfield{author}{\bibinfo{person}{Patrick~Gage Kelley},
  \bibinfo{person}{Lucian Cesca}, \bibinfo{person}{Joanna Bresee}, {and}
  \bibinfo{person}{Lorrie~Faith Cranor}.} \bibinfo{year}{2010}\natexlab{}.
\newblock \showarticletitle{Standardizing Privacy Notices: An Online Study of
  the Nutrition Label Approach}. In \bibinfo{booktitle}{\emph{Proceedings of
  the SIGCHI Conference on Human Factors in Computing Systems}} (Atlanta,
  Georgia, USA) \emph{(\bibinfo{series}{CHI '10})}.
  \bibinfo{publisher}{Association for Computing Machinery},
  \bibinfo{address}{New York, NY, USA}, \bibinfo{pages}{1573–1582}.
\newblock
\showISBNx{9781605589299}
\urldef\tempurl%
\url{https://doi.org/10.1145/1753326.1753561}
\showDOI{\tempurl}


\bibitem[\protect\citeauthoryear{Kesler}{Kesler}{2022}]%
        {kesler_att_2022}
\bibfield{author}{\bibinfo{person}{Reinhold Kesler}.}
  \bibinfo{year}{2022}\natexlab{}.
\newblock \showarticletitle{The Impact of Apple’s App Tracking Transparency
  on App Monetization}.
\newblock \bibinfo{journal}{\emph{Work in Progress}} (\bibinfo{year}{2022}),
  \bibinfo{numpages}{22}~pages.
\newblock


\bibitem[\protect\citeauthoryear{Kollnig}{Kollnig}{2019}]%
        {kollnig_tracking_2019}
\bibfield{author}{\bibinfo{person}{Konrad Kollnig}.}
  \bibinfo{year}{2019}\natexlab{}.
\newblock \showarticletitle{Tracking in Apps' Privacy Policies}.
\newblock \bibinfo{journal}{\emph{arXiv preprint arXiv:2111.07860}}
  (\bibinfo{year}{2019}), \bibinfo{numpages}{10}~pages.
\newblock
\showeprint[arxiv]{2111.07860}~[cs]
\urldef\tempurl%
\url{http://arxiv.org/abs/2111.07860}
\showURL{%
\tempurl}


\bibitem[\protect\citeauthoryear{Kollnig, Binns, Dewitte, {Van Kleek}, Wang,
  Omeiza, Webb, and Shadbolt}{Kollnig et~al\mbox{.}}{2021a}]%
        {kollnig_2021}
\bibfield{author}{\bibinfo{person}{Konrad Kollnig}, \bibinfo{person}{Reuben
  Binns}, \bibinfo{person}{Pierre Dewitte}, \bibinfo{person}{Max {Van Kleek}},
  \bibinfo{person}{Ge Wang}, \bibinfo{person}{Daniel Omeiza},
  \bibinfo{person}{Helena Webb}, {and} \bibinfo{person}{Nigel Shadbolt}.}
  \bibinfo{year}{2021}\natexlab{a}.
\newblock \showarticletitle{A Fait Accompli? An Empirical~Study into the
  Absence of Consent to Third-Party Tracking in Android Apps}.
\newblock \bibinfo{journal}{\emph{Proceedings of the Seventeenth Symposium on
  Usable Privacy and Security}} (\bibinfo{year}{2021}).
\newblock


\bibitem[\protect\citeauthoryear{Kollnig, Binns, Van~Kleek, Lyngs, Zhao,
  Tinsman, and Shadbolt}{Kollnig et~al\mbox{.}}{2021b}]%
        {kollnig_before_2021}
\bibfield{author}{\bibinfo{person}{Konrad Kollnig}, \bibinfo{person}{Reuben
  Binns}, \bibinfo{person}{Max Van~Kleek}, \bibinfo{person}{Ulrik Lyngs},
  \bibinfo{person}{Jun Zhao}, \bibinfo{person}{Claudine Tinsman}, {and}
  \bibinfo{person}{Nigel Shadbolt}.} \bibinfo{year}{2021}\natexlab{b}.
\newblock \showarticletitle{Before and after {GDPR}: Tracking in Mobile Apps}.
\newblock  \bibinfo{volume}{10}, \bibinfo{number}{4} (\bibinfo{year}{2021}),
  \bibinfo{numpages}{30}~pages.
\newblock
\showISSN{2197-6775}
\urldef\tempurl%
\url{https://doi.org/10.14763/2021.4.1611}
\showDOI{\tempurl}


\bibitem[\protect\citeauthoryear{Kollnig, Shuba, Binns, Kleek, and
  Shadbolt}{Kollnig et~al\mbox{.}}{2022}]%
        {kollnig2021iphones}
\bibfield{author}{\bibinfo{person}{Konrad Kollnig}, \bibinfo{person}{Anastasia
  Shuba}, \bibinfo{person}{Reuben Binns}, \bibinfo{person}{Max~Van Kleek},
  {and} \bibinfo{person}{Nigel Shadbolt}.} \bibinfo{year}{2022}\natexlab{}.
\newblock \showarticletitle{Are {iPhones} Really Better for Privacy? A
  Comparative Study of {iOS} and {Android} Apps}.
\newblock \bibinfo{journal}{\emph{Proceedings on Privacy Enhancing
  Technologies}} \bibinfo{volume}{2022}, \bibinfo{number}{2}
  (\bibinfo{year}{2022}), \bibinfo{pages}{6--24}.
\newblock
\urldef\tempurl%
\url{https://doi.org/10.2478/popets-2022-0033}
\showDOI{\tempurl}


\bibitem[\protect\citeauthoryear{Leith}{Leith}{2021}]%
        {leith_mobile_2021}
\bibfield{author}{\bibinfo{person}{Douglas~J Leith}.}
  \bibinfo{year}{2021}\natexlab{}.
\newblock \showarticletitle{Mobile {{Handset Privacy}}: {{Measuring The Data
  iOS}} and {{Android Send}} to {{Apple And Google}}}.
\newblock  (\bibinfo{year}{2021}), \bibinfo{pages}{10}.
\newblock


\bibitem[\protect\citeauthoryear{{Lockdown Privacy}}{{Lockdown
  Privacy}}{2021}]%
        {apple_enforcement2}
\bibfield{author}{\bibinfo{person}{{Lockdown Privacy}}.}
  \bibinfo{year}{2021}\natexlab{}.
\newblock \bibinfo{title}{{Study: Effectiveness of Apple's App Tracking
  Transparency}}.
\newblock
  \bibinfo{howpublished}{\url{https://blog.lockdownprivacy.com/2021/09/22/study-effectiveness-of-apples-app-tracking-transparency.html}}.
\newblock


\bibitem[\protect\citeauthoryear{McDonald and Cranor}{McDonald and
  Cranor}{2008}]%
        {mcdonald_cost_2008}
\bibfield{author}{\bibinfo{person}{Aleecia~M McDonald} {and}
  \bibinfo{person}{Lorrie~Faith Cranor}.} \bibinfo{year}{2008}\natexlab{}.
\newblock \showarticletitle{The {{Cost}} of {{Reading Privacy Policies}}}.
\newblock \bibinfo{journal}{\emph{I/S: A Journal of Law and Policy for the
  Information Society}} (\bibinfo{year}{2008}), \bibinfo{pages}{26}.
\newblock


\bibitem[\protect\citeauthoryear{Mhaidli, Zou, and Schaub}{Mhaidli
  et~al\mbox{.}}{2019}]%
        {mhaidli_we_2019}
\bibfield{author}{\bibinfo{person}{Abraham~H Mhaidli}, \bibinfo{person}{Yixin
  Zou}, {and} \bibinfo{person}{Florian Schaub}.}
  \bibinfo{year}{2019}\natexlab{}.
\newblock \showarticletitle{``{{We Can}}'t {{Live Without Them}}!'' {{App
  Developers}}' {{Adoption}} of {{Ad Networks}} and {{Their Considerations}} of
  {{Consumer Risks}}}.
\newblock \bibinfo{journal}{\emph{Proceedings of the Fifteenth Symposium on
  Usable Privacy and Security}} (\bibinfo{year}{2019}), \bibinfo{pages}{21}.
\newblock


\bibitem[\protect\citeauthoryear{{Mobile Dev Memo}}{{Mobile Dev Memo}}{2021a}]%
        {apple_selfpreference}
\bibfield{author}{\bibinfo{person}{{Mobile Dev Memo}}.}
  \bibinfo{year}{2021}\natexlab{a}.
\newblock \bibinfo{title}{{ATT advantages Apple’s ad network. Here’s how to
  fix that.}}
\newblock
  \bibinfo{howpublished}{\url{https://mobiledevmemo.com/att-advantages-apples-ad-network-heres-how-to-fix-that/}}.
\newblock


\bibitem[\protect\citeauthoryear{{Mobile Dev Memo}}{{Mobile Dev Memo}}{2021b}]%
        {apple_enforcement1}
\bibfield{author}{\bibinfo{person}{{Mobile Dev Memo}}.}
  \bibinfo{year}{2021}\natexlab{b}.
\newblock \bibinfo{title}{{Why isn’t Apple policing mobile ads
  fingerprinting?}}
\newblock
  \bibinfo{howpublished}{\url{https://mobiledevmemo.com/why-isnt-apple-policing-mobile-ads-fingerprinting/}}.
\newblock


\bibitem[\protect\citeauthoryear{Nguyen, Backes, Marnau, and Stock}{Nguyen
  et~al\mbox{.}}{2021a}]%
        {nguyen_share_first_consent_2021}
\bibfield{author}{\bibinfo{person}{Trung~Tin Nguyen}, \bibinfo{person}{Michael
  Backes}, \bibinfo{person}{Ninja Marnau}, {and} \bibinfo{person}{Ben Stock}.}
  \bibinfo{year}{2021}\natexlab{a}.
\newblock \showarticletitle{Share First, Ask Later (or Never?) Studying
  Violations of GDPR{\textquoteright}s Explicit Consent in Android Apps}. In
  \bibinfo{booktitle}{\emph{30th {USENIX} Security Symposium ({USENIX} Security
  21)}}. \bibinfo{publisher}{{USENIX} Association},
  \bibinfo{pages}{3667--3684}.
\newblock
\showISBNx{978-1-939133-24-3}
\urldef\tempurl%
\url{https://www.usenix.org/conference/usenixsecurity21/presentation/nguyen}
\showURL{%
\tempurl}


\bibitem[\protect\citeauthoryear{Nguyen, Backes, Marnau, and Stock}{Nguyen
  et~al\mbox{.}}{2021b}]%
        {apps_consent_2021}
\bibfield{author}{\bibinfo{person}{Trung~Tin Nguyen}, \bibinfo{person}{Michael
  Backes}, \bibinfo{person}{Ninja Marnau}, {and} \bibinfo{person}{Ben Stock}.}
  \bibinfo{year}{2021}\natexlab{b}.
\newblock \showarticletitle{Share First, Ask Later (or Never?) Studying
  Violations of {GDPR{\textquoteright}s} Explicit Consent in Android Apps}. In
  \bibinfo{booktitle}{\emph{30th USENIX Security Symposium (USENIX Security
  21)}}. \bibinfo{publisher}{USENIX Association}, \bibinfo{pages}{3667--3684}.
\newblock
\showISBNx{978-1-939133-24-3}
\urldef\tempurl%
\url{https://www.usenix.org/conference/usenixsecurity21/presentation/nguyen}
\showURL{%
\tempurl}


\bibitem[\protect\citeauthoryear{Nissenbaum}{Nissenbaum}{2004}]%
        {nissenbaum_privacy_2004}
\bibfield{author}{\bibinfo{person}{Helen Nissenbaum}.}
  \bibinfo{year}{2004}\natexlab{}.
\newblock \showarticletitle{Privacy as Contextual Integrity}.
\newblock \bibinfo{journal}{\emph{Washington Law Review}}  \bibinfo{volume}{79}
  (\bibinfo{year}{2004}), \bibinfo{pages}{39}.
\newblock


\bibitem[\protect\citeauthoryear{Okoyomon, Samarin, Wijesekera, Elazari,
  Vallina-Rodriguez, Reyes, Feal, and Egelman}{Okoyomon et~al\mbox{.}}{2019}]%
        {okoyomon_ridiculousness_2019}
\bibfield{author}{\bibinfo{person}{Ehimare Okoyomon}, \bibinfo{person}{Nikita
  Samarin}, \bibinfo{person}{Primal Wijesekera}, \bibinfo{person}{Amit
  Elazari}, \bibinfo{person}{Narseo Vallina-Rodriguez}, \bibinfo{person}{Irwin
  Reyes}, \bibinfo{person}{Alvaro Feal}, {and} \bibinfo{person}{Serge
  Egelman}.} \bibinfo{year}{2019}\natexlab{}.
\newblock \showarticletitle{On {{The Ridiculousness}} of {{Notice}} and
  {{Consent}}: {{Contradictions}} in {{App Privacy Policies}}}.
\newblock \bibinfo{journal}{\emph{The {{Workshop}} on {{Technology}} and
  {{Consumer Protection}} ({{ConPro}} ’19)}} (\bibinfo{year}{2019}),
  \bibinfo{numpages}{7}~pages.
\newblock


\bibitem[\protect\citeauthoryear{Ren, Rao, Lindorfer, Legout, and Choffnes}{Ren
  et~al\mbox{.}}{2016}]%
        {ren_recon_2016}
\bibfield{author}{\bibinfo{person}{Jingjing Ren}, \bibinfo{person}{Ashwin Rao},
  \bibinfo{person}{Martina Lindorfer}, \bibinfo{person}{Arnaud Legout}, {and}
  \bibinfo{person}{David Choffnes}.} \bibinfo{year}{2016}\natexlab{}.
\newblock \showarticletitle{{{ReCon}}: {{Revealing}} and {{Controlling PII
  Leaks}} in {{Mobile Network Traffic}}}. In
  \bibinfo{booktitle}{\emph{Proceedings of the 14th {{Annual International
  Conference}} on {{Mobile Systems}}, {{Applications}}, and {{Services}} -
  {{MobiSys}} '16}}. \bibinfo{publisher}{{ACM Press}},
  \bibinfo{address}{{Singapore, Singapore}}, \bibinfo{pages}{361--374}.
\newblock
\showISBNx{978-1-4503-4269-8}
\urldef\tempurl%
\url{https://doi.org/10.1145/2906388.2906392}
\showDOI{\tempurl}


\bibitem[\protect\citeauthoryear{{Reuters}}{{Reuters}}{uers}]%
        {korea}
\bibfield{author}{\bibinfo{person}{{Reuters}}.}
  \bibinfo{year}{Reteuers}\natexlab{}.
\newblock \bibinfo{title}{{S.Korea targets Apple over new app store
  regulation}}.
\newblock
  \bibinfo{howpublished}{\url{https://www.reuters.com/technology/skorea-targets-apple-over-new-app-store-regulation-2021-10-15/}}.
\newblock


\bibitem[\protect\citeauthoryear{Reyes, Wijesekera, Reardon, On, Razaghpanah,
  Vallina-Rodriguez, and Egelman}{Reyes et~al\mbox{.}}{2018}]%
        {reyes_wont_2018}
\bibfield{author}{\bibinfo{person}{Irwin Reyes}, \bibinfo{person}{Primal
  Wijesekera}, \bibinfo{person}{Joel Reardon}, \bibinfo{person}{Amit
  Elazari~Bar On}, \bibinfo{person}{Abbas Razaghpanah}, \bibinfo{person}{Narseo
  Vallina-Rodriguez}, {and} \bibinfo{person}{Serge Egelman}.}
  \bibinfo{year}{2018}\natexlab{}.
\newblock \showarticletitle{``{{Won}}'t {{Somebody Think}} of the
  {{Children}}?'' {{Examining COPPA Compliance}} at {{Scale}}}.
\newblock \bibinfo{journal}{\emph{Proceedings on Privacy Enhancing
  Technologies}} \bibinfo{volume}{2018}, \bibinfo{number}{3}
  (\bibinfo{year}{2018}), \bibinfo{pages}{63--83}.
\newblock
\showISSN{2299-0984}
\urldef\tempurl%
\url{https://doi.org/10.1515/popets-2018-0021}
\showDOI{\tempurl}


\bibitem[\protect\citeauthoryear{Shklovski, Mainwaring, Skúladóttir, and
  Borgthorsson}{Shklovski et~al\mbox{.}}{2014}]%
        {shklovski_leakiness_2014}
\bibfield{author}{\bibinfo{person}{Irina Shklovski}, \bibinfo{person}{Scott~D.
  Mainwaring}, \bibinfo{person}{Halla~Hrund Skúladóttir}, {and}
  \bibinfo{person}{Höskuldur Borgthorsson}.} \bibinfo{year}{2014}\natexlab{}.
\newblock \showarticletitle{Leakiness and Creepiness in App Space: Perceptions
  of Privacy and Mobile App Use}. In \bibinfo{booktitle}{\emph{Proceedings of
  the 32nd Annual {{ACM}} Conference on {{Human}} Factors in Computing Systems
  - {{CHI}} '14}} ({Toronto, Ontario, Canada}). \bibinfo{publisher}{{ACM
  Press}}, \bibinfo{pages}{2347--2356}.
\newblock
\showISBNx{978-1-4503-2473-1}
\urldef\tempurl%
\url{https://doi.org/10.1145/2556288.2557421}
\showDOI{\tempurl}


\bibitem[\protect\citeauthoryear{Shuba and Markopoulou}{Shuba and
  Markopoulou}{2020}]%
        {shuba_nomoats_2020}
\bibfield{author}{\bibinfo{person}{Anastasia Shuba} {and}
  \bibinfo{person}{Athina Markopoulou}.} \bibinfo{year}{2020}\natexlab{}.
\newblock \showarticletitle{{{NoMoATS}}: {{Towards Automatic Detection}} of
  {{Mobile Tracking}}}.
\newblock \bibinfo{journal}{\emph{Proceedings on Privacy Enhancing
  Technologies}} \bibinfo{volume}{2020}, \bibinfo{number}{2}
  (\bibinfo{year}{2020}), \bibinfo{pages}{45--66}.
\newblock
\showISSN{2299-0984}
\urldef\tempurl%
\url{https://doi.org/10.2478/popets-2020-0017}
\showDOI{\tempurl}


\bibitem[\protect\citeauthoryear{Shuba, Markopoulou, and Shafiq}{Shuba
  et~al\mbox{.}}{2018}]%
        {nomoads_2018}
\bibfield{author}{\bibinfo{person}{Anastasia Shuba}, \bibinfo{person}{Athina
  Markopoulou}, {and} \bibinfo{person}{Zubair Shafiq}.}
  \bibinfo{year}{2018}\natexlab{}.
\newblock \showarticletitle{NoMoAds: Effective and Efficient Cross-App Mobile
  Ad-Blocking}. In \bibinfo{booktitle}{\emph{Proceedings on Privacy Enhancing
  Technologies 2018}}. \bibinfo{pages}{125--140}.
\newblock
\urldef\tempurl%
\url{https://doi.org/10.1515/popets-2018-0035}
\showDOI{\tempurl}


\bibitem[\protect\citeauthoryear{Song and Hengartner}{Song and
  Hengartner}{2015}]%
        {privacyguard_vpn_2015}
\bibfield{author}{\bibinfo{person}{Yihang Song} {and} \bibinfo{person}{Urs
  Hengartner}.} \bibinfo{year}{2015}\natexlab{}.
\newblock \showarticletitle{{{PrivacyGuard}}: A {{VPN}}-based Platform to
  Detect Information Leakage on Android Devices}. In
  \bibinfo{booktitle}{\emph{Proceedings of the 5th Annual ACM CCS Workshop on
  Security and Privacy in Smartphones and Mobile Devices}}
  \emph{(\bibinfo{series}{SPSM '15})}. \bibinfo{pages}{15--26}.
\newblock
\urldef\tempurl%
\url{https://doi.org/10.1145/2808117.2808120}
\showDOI{\tempurl}


\bibitem[\protect\citeauthoryear{{The Verge}}{{The Verge}}{2013}]%
        {uuid_deprecation}
\bibfield{author}{\bibinfo{person}{{The Verge}}.}
  \bibinfo{year}{2013}\natexlab{}.
\newblock \bibinfo{title}{{Apple to finally stop accepting apps that use
  outdated UDID device identifier on May 1st}}.
\newblock
  \bibinfo{howpublished}{\url{https://www.theverge.com/2013/3/21/4133288/apple-to-finally-stop-accepting-apps-that-use-outdated-udid-device-identifier-may-1st}}.
\newblock


\bibitem[\protect\citeauthoryear{{van Hoboken} and Fathaigh}{{van Hoboken} and
  Fathaigh}{2021}]%
        {hoboken2021}
\bibfield{author}{\bibinfo{person}{Joris {van Hoboken}} {and}
  \bibinfo{person}{R~O Fathaigh}.} \bibinfo{year}{2021}\natexlab{}.
\newblock \showarticletitle{Smartphone platforms as privacy regulators}.
\newblock \bibinfo{journal}{\emph{Computer Law \& Security Review}}
  \bibinfo{volume}{41} (\bibinfo{year}{2021}), \bibinfo{pages}{105557}.
\newblock
\showISSN{0267-3649}
\urldef\tempurl%
\url{https://doi.org/10.1016/j.clsr.2021.105557}
\showDOI{\tempurl}


\bibitem[\protect\citeauthoryear{Van~Kleek, Binns, Zhao, Slack, Lee, Ottewell,
  and Shadbolt}{Van~Kleek et~al\mbox{.}}{2018}]%
        {van_kleek_x-ray_2018}
\bibfield{author}{\bibinfo{person}{Max Van~Kleek}, \bibinfo{person}{Reuben
  Binns}, \bibinfo{person}{Jun Zhao}, \bibinfo{person}{Adam Slack},
  \bibinfo{person}{Sauyon Lee}, \bibinfo{person}{Dean Ottewell}, {and}
  \bibinfo{person}{Nigel Shadbolt}.} \bibinfo{year}{2018}\natexlab{}.
\newblock \showarticletitle{X-{{Ray Refine}}: {{Supporting}} the
  {{Exploration}} and {{Refinement}} of {{Information Exposure Resulting}} from
  {{Smartphone Apps}}}. In \bibinfo{booktitle}{\emph{Proceedings of the 2018
  {{CHI Conference}} on {{Human Factors}} in {{Computing Systems}} - {{CHI}}
  '18}} ({Montreal QC, Canada}). \bibinfo{publisher}{{ACM Press}},
  \bibinfo{pages}{1--13}.
\newblock
\showISBNx{978-1-4503-5620-6}
\urldef\tempurl%
\url{https://doi.org/10.1145/3173574.3173967}
\showDOI{\tempurl}


\bibitem[\protect\citeauthoryear{{Van Kleek}, Liccardi, Binns, Zhao, Weitzner,
  and Shadbolt}{{Van Kleek} et~al\mbox{.}}{2017}]%
        {van_kleek_better_2017}
\bibfield{author}{\bibinfo{person}{Max {Van Kleek}}, \bibinfo{person}{Ilaria
  Liccardi}, \bibinfo{person}{Reuben Binns}, \bibinfo{person}{Jun Zhao},
  \bibinfo{person}{Daniel~J. Weitzner}, {and} \bibinfo{person}{Nigel
  Shadbolt}.} \bibinfo{year}{2017}\natexlab{}.
\newblock \showarticletitle{Better the {{Devil You Know}}: {{Exposing}} the
  {{Data Sharing Practices}} of {{Smartphone Apps}}}. In
  \bibinfo{booktitle}{\emph{Proceedings of the 2017 {{CHI Conference}} on
  {{Human Factors}} in {{Computing Systems}} - {{CHI}} '17}}.
  \bibinfo{publisher}{{ACM Press}}, \bibinfo{pages}{5208--5220}.
\newblock
\showISBNx{978-1-4503-4655-9}
\urldef\tempurl%
\url{https://doi.org/10.1145/3025453.3025556}
\showDOI{\tempurl}


\bibitem[\protect\citeauthoryear{Viennot, Garcia, and Nieh}{Viennot
  et~al\mbox{.}}{2014}]%
        {playdrone_2014}
\bibfield{author}{\bibinfo{person}{Nicolas Viennot}, \bibinfo{person}{Edward
  Garcia}, {and} \bibinfo{person}{Jason Nieh}.}
  \bibinfo{year}{2014}\natexlab{}.
\newblock \showarticletitle{A Measurement Study of {{Google Play}}}. In
  \bibinfo{booktitle}{\emph{The 2014 ACM International Conference on
  Measurement and Modeling of Computer Systems}}
  \emph{(\bibinfo{series}{SIGMETRICS '14})}. \bibinfo{pages}{221--233}.
\newblock
\urldef\tempurl%
\url{https://doi.org/10.1145/2591971.2592003}
\showDOI{\tempurl}


\bibitem[\protect\citeauthoryear{{W3C Working Group}}{{W3C Working
  Group}}{2019}]%
        {w3c_tracking_definition}
\bibfield{author}{\bibinfo{person}{{W3C Working Group}}.}
  \bibinfo{year}{2019}\natexlab{}.
\newblock \bibinfo{title}{{Tracking Compliance and Scope}}.
\newblock
  \bibinfo{howpublished}{\url{https://www.w3.org/TR/tracking-compliance/\#tracking}}.
\newblock


\bibitem[\protect\citeauthoryear{Wang, Liu, Liang, Vallina-Rodriguez, Guo, Li,
  Tapiador, Cao, and Xu}{Wang et~al\mbox{.}}{2018}]%
        {china_2018}
\bibfield{author}{\bibinfo{person}{Haoyu Wang}, \bibinfo{person}{Zhe Liu},
  \bibinfo{person}{Jingyue Liang}, \bibinfo{person}{Narseo Vallina-Rodriguez},
  \bibinfo{person}{Yao Guo}, \bibinfo{person}{Li Li}, \bibinfo{person}{Juan
  Tapiador}, \bibinfo{person}{Jingcun Cao}, {and} \bibinfo{person}{Guoai Xu}.}
  \bibinfo{year}{2018}\natexlab{}.
\newblock \showarticletitle{Beyond {{Google Play}}: A Large-Scale Comparative
  Study of {{Chinese}} Android App Markets}. In
  \bibinfo{booktitle}{\emph{Proceedings of the Internet Measurement Conference
  2018}} \emph{(\bibinfo{series}{IMC '18})}. \bibinfo{pages}{293--307}.
\newblock
\urldef\tempurl%
\url{https://doi.org/10.1145/3278532.3278558}
\showDOI{\tempurl}


\bibitem[\protect\citeauthoryear{{Washington Post}}{{Washington Post}}{2021a}]%
        {nutrition_labels}
\bibfield{author}{\bibinfo{person}{{Washington Post}}.}
  \bibinfo{year}{2021}\natexlab{a}.
\newblock \bibinfo{title}{{I checked Apple’s new privacy ‘nutrition
  labels.’ Many were false.}}
\newblock
  \bibinfo{howpublished}{\url{https://www.washingtonpost.com/technology/2021/01/29/apple-privacy-nutrition-label/}}.
\newblock


\bibitem[\protect\citeauthoryear{{Washington Post}}{{Washington Post}}{2021b}]%
        {apple_enforcement3}
\bibfield{author}{\bibinfo{person}{{Washington Post}}.}
  \bibinfo{year}{2021}\natexlab{b}.
\newblock \bibinfo{title}{{When you ‘Ask app not to track,’ some iPhone
  apps keep snooping anyway}}.
\newblock
  \bibinfo{howpublished}{\url{https://www.washingtonpost.com/technology/2021/09/23/iphone-tracking/}}.
\newblock


\bibitem[\protect\citeauthoryear{Zimmeck, Story, Smullen, Ravichander, Wang,
  Reidenberg, Russell, and Sadeh}{Zimmeck et~al\mbox{.}}{2019}]%
        {maps_2019}
\bibfield{author}{\bibinfo{person}{Sebastian Zimmeck}, \bibinfo{person}{Peter
  Story}, \bibinfo{person}{Daniel Smullen}, \bibinfo{person}{Abhilasha
  Ravichander}, \bibinfo{person}{Ziqi Wang}, \bibinfo{person}{Joel Reidenberg},
  \bibinfo{person}{N.~Cameron Russell}, {and} \bibinfo{person}{Norman Sadeh}.}
  \bibinfo{year}{2019}\natexlab{}.
\newblock \showarticletitle{{MAPS}: Scaling Privacy Compliance Analysis to a
  Million Apps}.
\newblock \bibinfo{journal}{\emph{Proceedings on Privacy Enhancing
  Technologies}} \bibinfo{volume}{2019}, \bibinfo{number}{3}
  (\bibinfo{year}{2019}), \bibinfo{pages}{66--86}.
\newblock
\urldef\tempurl%
\url{https://doi.org/10.2478/popets-2019-0037}
\showDOI{\tempurl}


\bibitem[\protect\citeauthoryear{Ó~Fathaigh and van Hoboken}{Ó~Fathaigh and
  van Hoboken}{2019}]%
        {o_fathaigh_european_2019}
\bibfield{author}{\bibinfo{person}{R. Ó~Fathaigh} {and} \bibinfo{person}{J.
  van Hoboken}.} \bibinfo{year}{2019}\natexlab{}.
\newblock \showarticletitle{European {{Regulation}} of {{Smartphone
  Ecosystems}}}.
\newblock \bibinfo{journal}{\emph{European Data Protection Law Review}}
  \bibinfo{volume}{5}, \bibinfo{number}{4} (\bibinfo{year}{2019}),
  \bibinfo{pages}{476--491}.
\newblock
\showISSN{23642831, 2364284X}
\urldef\tempurl%
\url{https://doi.org/10.21552/edpl/2019/4/6}
\showDOI{\tempurl}


\end{thebibliography}
